\documentclass[aps,pre,twocolumn,groupedaddress,amsmath,showpacs,eqsecnum]{revtex4}

   \usepackage{bm}
\usepackage{graphicx,amssymb}
\usepackage{bm}
\usepackage{amsmath}
\newcommand\cc{{\text{c}}}
\newcommand\kk{{\text{th}}}
\newcommand\bb{{\text{b}}}

\newcommand\zero{{(0)}}
\newcommand\beq{\begin{equation}}
\newcommand\eeq{\end{equation}}
\newcommand\beqa{\begin{eqnarray}}
\newcommand\eeqa{\end{eqnarray}}
\newcommand{\dd}{\text{d}}
\newcommand{\e}{\text{e}}
\begin{document}

\title{Granular fluid thermostatted by a bath of elastic hard spheres}
\author{Andr\'es Santos}
\email{andres@unex.es}
 \homepage{http://www.unex.es/fisteor/andres/}
\affiliation{Departamento de F\'{\i}sica, Universidad de Extremadura,
E--06071 Badajoz, Spain}

\date{\today}
\begin{abstract}
The homogeneous steady state of a fluid of inelastic hard spheres immersed in a bath of elastic hard spheres kept at equilibrium is analyzed by means of the first Sonine approximation to the (spatially homogeneous) Enskog--Boltzmann equation. The temperature of the granular fluid relative to the bath temperature and the kurtosis of the granular distribution function are obtained as functions of the coefficient of restitution, the mass ratio, and a dimensionless parameter $\beta$ measuring the cooling rate relative to the friction constant. Comparison with recent results obtained from an iterative numerical solution of the Enskog--Boltzmann equation [Biben et al., Physica A \textbf{310}, 308 (202)] shows an excellent agreement. Several limiting cases are also considered. In particular, when the granular particles are much heavier than the bath particles (but have a comparable size and number density), it is shown that the bath acts as a white noise external driving. 
In the general case, 
the Sonine approximation predicts the lack of a steady state if the control parameter $\beta$ is larger than a certain critical value $\beta_c$ that depends on the coefficient of restitution and the mass ratio. However, this phenomenon  appears outside the expected domain of applicability of the approximation. 
\end{abstract}
\pacs{45.70.-n, 05.20.Dd, 51.10.+y}

\maketitle
\section{Introduction\label{sec1}}
The simplest model to describe the dynamics of granular matter in the regime of 
rapid flow consists of an assembly of (smooth) \textit{inelastic} hard spheres  
with a constant coefficient of normal restitution $\alpha$ \cite{C90}.
The Liouville operator and the BBGKY hierarchy governing the time evolution of the phase space density and the reduced distribution functions, respectively, can be extended to the case of dissipative collisions \cite{BDS97,vNEB98}. By assuming that the pre-collision velocities of two particles at contact are uncorrelated (molecular chaos assumption), an approximate kinetic equation for the one-body velocity distribution function can be derived, thus extending the revised Enskog theory to the realm of dissipative dynamics \cite{BDS97,DBS97}.
The Enskog equation accounts for spatial correlations through the equilibrium pair correlation function at contact $g[n(\mathbf{r})]$ as a functional of the nonequilibrium density field $n(\mathbf{r})$. On the other hand, in the special case of spatially uniform states the pair correlation function $g(n)$ becomes a constant, so that the Enskog equation reduces to the Boltzmann equation, except for an increase of the collision frequency proportional to $g(n)$.

The familiar concept of equilibrium is absent in a granular fluid due to the collisional dissipation of kinetic energy (which is transferred to the internal degrees of freedom) by an amount proportional to $1-\alpha^2$. Even if the system remains in a uniform state (the so-called homogeneous cooling state)
\cite{note0}, the total kinetic energy monotonically decreases with time, unless some kind of external forcing is acted upon the system to compensate for the collisional loss of energy and thus a steady state can be reached.
{}From an experimental point of view, a fluidized steady state is usually achieved by violently shaking the container \cite{OU98,LCDKG99,OU99,FWEFCGB99,RM00,BK01,WHP01,YHCMW02,WP02,FM02}. To mimic the effects of collisions with the vibrating walls in an otherwise \textit{uniform} system it has become popular to assume that each particle is subjected to a stochastic force with the properties of a Gaussian white noise \cite{WM96,W96,SBCM98,vNE98,vNETP99,BSSS99,MS00,PTvNE01}. The effect of this force is to produce frequent (and weak) random ``kicks'' to the particles between two successive collisions. On the other hand, the white noise force is not the only \textit{thermostatting} mechanism proposed in the literature \cite{CLH00}. For instance, energy can be injected into the system by the action of an ``anti-drag'' force (Gaussian thermostat) \cite{MS00,EB02} or a constant force directed along the motion direction (``gravity'' thermostat) \cite{MS00,EB02}. The former thermostat is equivalent to a velocity rescaling in the freely cooling state \cite{MS00}. 
On a different vein, Barrat et al.\ \cite{BTF01} have recently proposed a model in which the energy injected in vertically shaken granular systems is transferred to the horizontal degrees of freedom through collisions with an effective \textit{random} coefficient of restitution. 
Regardless of the heating mechanism, the common feature is that the granular fluid reaches a uniform nonequilibrium steady state characterized by a velocity distribution function $f(\mathbf{v})$ different from the equilibrium Maxwell--Boltzmann (MB) distribution
\beq
f_{\text{MB}}(\mathbf{v})=n\left(\frac{m}{2\pi T}\right)^{3/2}\e^{-mv^2/2T}.
\label{a1}
\eeq
In this equation, $m$ is the mass of a particle, $n$ is the number density, and $T$ is the \textit{granular} temperature. The two latter quantities are defined in terms of the velocity distribution function as
\beq
n=\int \dd \mathbf{v}f(\mathbf{v}),
\label{a2}
\eeq
\beq
T=\frac{m}{3}\langle v^2\rangle,
\label{a3}
\eeq
where
\beq
\langle v^k\rangle=\frac{1}{n}\int \dd \mathbf{v}v^kf(\mathbf{v}).
\label{a4}
\eeq
The deviation of $f$ from $f_{\text{MB}}$ is usually monitored by the value of the fourth cumulant (or kurtosis) 
\beq
\kappa\equiv \frac{3}{5}\frac{\langle v^4\rangle}{\langle v^2\rangle^2}-1.
\label{a5}
\eeq
A negative (positive) value of $\kappa$ implies that the distribution is flatter (less flat) around the mean than the normal one and thus the distribution is said to be platykurtic (leptokurtic).
In the cases of the white noise and Gaussian thermostats,  the  distribution is platykurtic ($\kappa<0$) for $\alpha\gtrsim 0.7$, while it is leptokurtic ($\kappa>0$) for smaller values of the coefficient of restitution $\alpha$ \cite{vNE98,MS00,BRC96}. On the other hand, the distribution is always platykurtic when the system is heated with the gravity thermostat \cite{MS00}. As for the magnitude of $\kappa$, it is typically smaller with the white noise thermostat than with the Gaussian and gravity thermostats. In the random  coefficient of restitution approach the value of $\kappa$ depends on the probability distribution of $\alpha$ (through the moments $\langle \alpha\rangle$ and  $\langle \alpha^4\rangle$), but is otherwise positive definite \cite{BTF01}. Another important measure of departure from equilibrium  is the high energy tail of the distribution. The asymptotic behavior of the distribution function for large velocities is generally of the form $\ln f(\mathbf{v})\sim - v^a$, where $a=1$, $a=\frac{3}{2}$, and $a=2$ for the white noise, Gaussian, and gravity thermostats, respectively \cite{vNE98,MS00,BRC96,EP97}. In the case of the random coefficient of restitution, the values of the exponent $a$ seem to be highly dependent on the probability distribution of $\alpha$, ranging from $a\approx 0.8$ to $a\approx 2$ \cite{BTF01}.

The lack of an equilibrium state in granular fluids is especially evident in the case of a mixture. According to the principle of equipartition of energy, a mixture of two fluids at equilibrium share the same temperature ($T_1=T_2$). On the other hand, a binary granular mixture  in the homogeneous cooling state or in a nonequilibrium steady state exhibits two different granular temperatures \cite{WP02,FM02,GD99,BDS99,MP99,SD01a,SD01b,DG01,MG02a,BT02a,DHGD02,BT02b,GD02,MG02b,G02,MG02c,AL02}.
In general, the temperature ratio $T_1/T_2$ depends on the mass ratio, the diameter ratio, the three coefficients of restitution involved, the mole fraction, and the volume fraction \cite{GD99,DHGD02}. Typically, the largest influence on $T_1/T_2$ is due to the mass ratio, the heaviest component having the largest temperature \cite{WP02,FM02,DHGD02,AL02}. An extreme example of this breakdown of the energy equipartition occurs in the homogeneous cooling of an infinitely heavy impurity particle in a sea of inelastic particles: if the cooling rate of the granular fluid  is smaller than an effective impurity--fluid collision frequency, then the impurity/fluid temperature ratio diverges \cite{BDS99,SD01a}, an effect that can be characterized as a second-order nonequilibrium phase transition \cite{SD01a,SD01b}.

In a recent paper, Biben et al.\ \cite{BMP02} have proposed an interesting alternative way of uniformly heating a granular fluid to achieve a nonequilibrium steady state. The granular particles are assumed to be immersed in a bath of \textit{elastic} particles kept at \textit{equilibrium} at a certain temperature $T_\bb$. In the steady state, the energy loss due to inelastic collisions among the granular particles is balanced by the energy gain due to elastic collisions of the granular particles with the bath particles. The velocity distribution of the bath is given by the MB expression (\ref{a1}), except that $n$ must be replaced by $n_\bb$ (number density of the bath particles) and $m$ must be replaced by $m_\bb$ (mass of a bath particle). This system cannot be considered as a true binary mixture since the bath is assumed to be unaffected by the granular fluid. The role of the bath fluid is to provide a thermostatting mechanism for the granular fluid. However, this problem retains most of the basic features found in a mixture, such as the effect of mass disparity and the competition between different length and time scales. Although the bath is at equilibrium, the inelastic nature of the collisions among the granular particles prevent  the steady state to be the equilibrium one \cite{BMP02}. First, the equipartition of energy is broken since the granular temperature $T$ differs from the bath temperature $T_\bb$ (actually, $T<T_\bb$). Second, the kurtosis $\kappa$ of the granular velocity distribution is not zero.
Interestingly enough, the temperature ratio $T/T_\bb$ and the kurtosis $\kappa$ are not only functions of the coefficient of restitution $\alpha$, but also of the mass ratio $m/m_\bb$ and a dimensionless parameter $\omega$ measuring the mean free path associated with the granular--bath collisions, relative to the one associated with the granular--granular collisions. 
In Ref.~\cite{BMP02}, Biben et al.\ solved numerically the (steady-state) Enskog--Boltzmann equation by an iterative method and obtained $T/T_\bb$ and $\kappa$ for several choices of the parameters in the range $0\leq\alpha<1$, $m/m_\bb\geq 1$, $\omega\leq 1$. In addition, they observed that their numerical results were consistent with a Gaussian high energy tail, $\ln f(\mathbf{v})\sim -v^2$, although this asymptotic behavior was  reached for extremely large velocities only \cite{BMP02}. 

The same system as in Ref.\ \cite{BMP02} has been studied by Barrat and Trizac \cite{BT02a} by approximating the granular velocity distribution function $f(\mathbf{v})$ by its MB form (\ref{a1}). This allows one to obtain a closed cubic equation for the temperature ratio $T/T_\bb$. Comparison with the iterative numerical solution of Ref.\ \cite{BMP02} shows that the MB approximation provides a good estimate of $T/T_\bb$ for the values of $m/m_\bb\geq 1$ and $\omega\leq 1$ considered \cite{BT02a}. On the other hand, the MB approximation is obviously unable to estimate the kurtosis $\kappa$ of the distribution function. 

The aim of this paper is to revisit the granular fluid thermostatted by a bath of elastic hard spheres proposed by Biben et al.\ \cite{BMP02}. The main goal is to obtain (approximate) algebraic expressions allowing one to get $T/T_\bb$ and $\kappa$ in terms of $\alpha$, $m/m_\bb$, and $\omega$ from the Enskog--Boltzmann equation in the  (first) Sonine approximation. This approximation consists of expanding $f(\mathbf{v})/f_{\text{MB}}(\mathbf{v})-1$ in generalized Laguerre (or Sonine) polynomials and retaining the first term only, whose coefficient is the kurtosis $\kappa$. When this approximate form is inserted into the second- and fourth-order moment equations and terms nonlinear in $\kappa$ are neglected, one obtains a closed tenth-degree equation for the temperature ratio $T/T_\bb$ and an explicit expression of $\kappa$ in terms of  $T/T_\bb$.
A similar method has been used by van Noije and Ernst \cite{vNE98} to estimate $\kappa$ as a function of $\alpha$ in the case of a granular fluid in the freely cooling regime or heated by a white noise thermostat, the results showing a good agreement with computer simulations \cite{BRC96,MS00}. However, the situation in the case where the thermostatting mechanism is provided by collisions with elastic particles is much more complicated. In particular, the $\alpha$-dependence of $\kappa$ is strongly influenced by two independent control parameters, namely the mass ratio $m/m_\bb$ and the mean free path ratio $\omega$. 

The organization and main results of the paper are the following. The problem and notation are presented in Sec.\ \ref{sec2}. To gain some insight into the different competing scales of the problem, the MB approximation is also worked out in Sec.\ \ref{sec2}. This suggests the use of the ratio $\beta \propto \omega(1+m/m_\bb)^{1/2}(1-\alpha^2)$ between the cooling rate of the granular particles and the granular--bath collision rate (or friction constant) as a control parameter more convenient than the mean free path ratio $\omega$.

The Sonine approximation is carried out in Sec.\ \ref{sec3}. As said before, this approximation yields a tenth-degree equation for $T/T_\bb$. 
A useful approximate solution to this equation is found by expanding the solution around the MB approximation and neglecting  terms nonlinear in the deviation.
In order to get more explicit results, some limiting behaviors are considered, including the ``colloidal'' limit [$m/m_\bb\to\infty$, $\omega\propto (m/m_\bb)^{-1}\to 0$, $\beta\propto (m/m_\bb)^{-1/2}\to 0$] and the ``white noise'' limit [$m/m_\bb\to\infty$, $\omega=\text{finite}$, $\beta\propto (m/m_\bb)^{1/2}\to\infty$]. In the latter limit, one recovers the case of the white noise thermostat \cite{vNE98}. 
Section \ref{sec5} is devoted to a comparison with the numerical solution of Ref.\ \cite{BMP02} for $0\leq\alpha<1$, $m/m_\bb\geq 1$, $\omega\leq 1$. An excellent agreement is found, thus validating the reliability of the Sonine approximation, at least for $\omega\leq 1$.

The Fokker--Planck limit ($m/m_\bb\to\infty$) of the Enskog--Boltzmann equation is considered in Sec.\ \ref{sec4}. An analysis of the high energy tail shows that for asymptotically large velocities ($v\gg \sqrt{2T_\bb/m}$) the distribution function tends to a Gaussian, namely $\ln f\approx -mv^2/2T_\bb$. If the granular temperature is much smaller than the bath temperature ($T\ll T_\bb$), there exists an intermediate range of velocities $\sqrt{2T/m}\ll v\ll \sqrt{2T_\bb/m}$ where the distribution function has the form of a stretched exponential, $\ln f\sim -v^{3/2}$. In the white noise limit one has $T/T_\bb\to 0$ and so the Gaussian form is pushed out to infinity and the stretched exponential becomes the only observable asymptotic behavior.

 Section \ref{secnew} shows that the Sonine approximation predicts an interesting singular behavior. As $\beta$ (or, equivalently, $\omega$) is increased for fixed values of $\alpha$ and $m/m_\bb$, a \textit{critical} value $\beta_c(\alpha,m/m_\bb)$ is reached beyond which no steady-state solution exists. While for $\beta<\beta_c(\alpha,m/m_\bb)$ the temperature ratio $T/T_\bb$ takes a well-defined stationary value, it decreases with time ($T/T_\bb\to 0 $) for $\beta>\beta_c(\alpha,m/m_\bb)$. Therefore, the Sonine approximation predicts the existence of a (first-order) phase transition when the friction constant becomes small enough as compared with the cooling rate and so the bath is unable to thermostat the granular fluid. Since this phenomenon takes place outside the domain of applicability of the Sonine approximation, it is not possible to assert in this context whether it is an artifact of the approximation or not.
The paper ends with a summary and some concluding remarks in Sec.\ \ref{sec6}.

\section{Basic equations. Maxwell--Boltzmann approximation\label{sec2}}
Consider a fluid of inelastic, smooth hard spheres (species 1) of mass 
$m_1=m$, diameter $\sigma_1=\sigma$, and coefficient of normal restitution 
$\alpha_{11}=\alpha$. The number density is $n_1=n$. The granular fluid is 
kept at a stationary, homogeneous, and isotropic state by a 
thermostat modeled as a bath made of a 
number density $n_2=n_\bb$ of hard spheres of mass $m_2=m_\bb$ and diameter 
$\sigma_2=\sigma_\bb$ (species 2). The granular particles are assumed to 
collide \textit{elastically}  with the bath particles (i.e.\ 
$\alpha_{12}=1$). In addition, the bath is assumed to be at equilibrium, 
unaffected by the granular fluid. Its velocity distribution function 
$f_2=f_\bb$ is then
\beq
f_\bb(\mathbf{v})=n_\bb \left(\frac{m_\bb}{2\pi T_\bb}\right)^{3/2} \exp\left(-\frac{m_\bb 
v^2}{2T_\bb}\right).
\label{1}
\eeq

Assuming the validity of the Enskog--Boltzmann description, the velocity distribution function $f_1=f$ of the granular 
fluid satisfies the equation
\beq
\partial_t f_1(\mathbf{v})=J_{11}[\mathbf{v}_1|f_1,f_1]+J_{12}[\mathbf{v}_1|f_1,f_2].
\label{2}
\eeq
The collision 
operators $J_{ij}[\mathbf{v}_1|f_i,f_j]$ are \cite{GD99}
\beqa
J_{ij}[\mathbf{v}_1|f_i,f_j]&=&g_{ij}\sigma_{ij}^2\int \dd\mathbf{v}_2\int 
\dd\widehat{\bm{\sigma}}\,\Theta(\widehat{\bm{\sigma}}\cdot\mathbf{v}_{12})
(\widehat{\bm{\sigma}}\cdot\mathbf{v}_{12}) 
\nonumber\\
&&\times\left[\alpha_{ij}^{-2}f_i(\mathbf{v}_1')f_j(\mathbf{v}_2')-f_i(\mathbf{v}_1)f_j(\mathbf{v}_2)\right],\nonumber\\
&&
\label{3}
\eeqa
where $g_{ij}$ is the pair correlation function for particles of species $i$ and $j$ 
at the contact point $r=\sigma_{ij}\equiv (\sigma_i+\sigma_j)/2$. The 
primes denote pre-collisional velocities,
\begin{subequations}
\beq
\mathbf{v}_1'=\mathbf{v}_1-\mu_{ji}\left(1+\alpha_{ij}^{-1}\right)
(\widehat{\bm{\sigma}}\cdot\mathbf{v}_{12})\widehat{\bm{\sigma}},
\eeq
\beq
\mathbf{v}_2'=\mathbf{v}_2+\mu_{ij}\left(1+\alpha_{ij}^{-1}\right)
(\widehat{\bm{\sigma}}\cdot\mathbf{v}_{12})\widehat{\bm{\sigma}},
\eeq
\label{4}
\end{subequations}
where $\mathbf{v}_{12}\equiv\mathbf{v}_1-\mathbf{v}_2$ and $\mu_{ij}\equiv 
m_i/(m_i+m_j)$.
The first term on the right-hand side of Eq.\ (\ref{2}) is a nonlinear collision operator describing the rate of change of $f$ due to the \textit{inelastic} collisions among granular particles, while the second term is a linear collision operator (Boltzmann--Lorentz operator) accounting for the \textit{elastic} collisions of granular particles against bath particles.
Since mass is conserved in a single collision, we have the identity
\beq
\int \dd\mathbf{v}_1\,  J_{ij}[\mathbf{v}_1|f_i,f_j]=0.
\label{b1}
\eeq
On the other hand, energy is not conserved by collisions. Multiplying both sides of Eq.\ (\ref{2}) by $v_1^2$ and integrating over velocity we get
\beq
\partial_t\langle v^2\rangle=-(A_{11}+A_{12}),
\label{8bis}
\eeq
where
\beq
A_{ij}\equiv-\frac{1}{n}\int  \dd\mathbf{v}_1\, v_1^2 J_{ij}[\mathbf{v}_1|f_i,f_j].
\label{10}
\eeq
The inelastic 1--1 collisions produce an energy loss, i.e.\ $A_{11}>0$. As a matter of fact,
the quantity $A_{11}$ is proportional to the cooling rate of the 
granular fluid $\zeta$:
\beq
\zeta=\frac{A_{11}}{3T/m}.
\label{10bis}
\eeq
 Collisions 1--2 are elastic, so the total kinetic energy of the two colliding partners is preserved. Therefore, the sign of $A_{12}$ depends on whether $T>T_\bb$ or $T<T_\bb$. Since in the steady state the mean kinetic energy of the granular particles is smaller than that of the bath particles ($T<T_\bb$), the granular partner gains energy on average, so $A_{12}<0$. In the \textit{steady state} both terms exactly compensate each other and one has
\beq
A_{11}+A_{12}=0.
\label{8}
\eeq
Analogously,
\beq
\partial_t\langle v^4\rangle=-(B_{11}+B_{12}),
\label{9bis}
\eeq
where  $B_{ij}$ are the collisional moments
\beq
B_{ij}=-\frac{1}{n}\int \dd\mathbf{v}_1\, v_1^4 J_{ij}[\mathbf{v}_1|f_i,f_j].
\label{11}
\eeq
In the steady state,
\beq
B_{11}+B_{12}=0.
\label{9}
\eeq

The  main objective of this paper is to determine the granular temperature $T_1=T$ and the 
fourth cumulant $\kappa$, which are related to the second and fourth moments 
of the distribution function by Eqs.\ (\ref{a3}) and (\ref{a5}).
To that end, 
I will use a first Sonine approximation in the next Section. 
Before that, it is instructive to get an estimate of $T/T_\bb$ by means of the much simpler MB approximation for $f$, as done in Ref.~\cite{BT02a}. 
This also will allow us to introduce the two competing rates, along with some useful dimensionless parameters.
If $f$ is replaced by $f_{\text{MB}}$ [cf.\ Eq.\ (\ref{a1})] in Eq.\ (\ref{10}), one gets
\beq
A_{ij}\approx A_{ij}^\zero,
\label{b2}
\eeq
where \cite{vNE98,GD99}
\beq
A_{11}^{(0)}=\frac{3}{2}\nu\frac{2T}{m}(1-\alpha^2),
\label{14}
\eeq
\beq
A_{12}^{(0)}=3\gamma\frac{2T}{m}
x(1-x^2).
\label{32}
\eeq
In Eq.\ (\ref{14}),
\beq
\nu=\frac{2}{3}\sqrt{2\pi}\left(\frac{2T}{m}\right)^{1/2}n\sigma^2g_{11}
\label{17.1}
\eeq
is an effective (self) collision frequency.  In terms of it, the cooling rate  (\ref{10bis}) is approximated by $\zeta\approx \zeta^\zero$, where
\beq	
\zeta^\zero=\nu(1-\alpha^2).
\label{17.2}
\eeq
In Eq.\ (\ref{32}), 
\beq
x\equiv \left[\mu+\frac{T_\bb}{T}(1-\mu)\right]^{1/2}
\label{36}
\eeq
 is a parameter measuring the temperature ratio, where $\mu\equiv \mu_{21}=(1+m/m_\bb)^{-1}$, and
\beq 
\gamma=\frac{8\sqrt{\pi}}{3}n_\bb\sigma_{12}^2  
g_{12}\mu^{1/2}\left(\frac{2T}{m}\right)^{1/2}
\label{19}
\eeq
defines a characteristic rate for (cross) collisions of the granular particles with the bath particles. In the limit of one thermalized Brownian particle  ($n\to 0$, $m/m_\bb\to \infty$, $T=T_\bb$), it is easy to show that $\partial_t\langle\mathbf{v}\rangle=-\gamma\langle\mathbf{v}\rangle$, so $\gamma$ plays the role of a \textit{friction} constant. Henceforth I will use this terminology when referring to $\gamma$. The cooling rate $\zeta^{(0)}$ and the friction constant  $\gamma$ define the two relevant time scales of the problem. Their ratio is
\beq
\beta\equiv \frac{\zeta^\zero}{\gamma}=2^{-3/2}\omega{\mu}^{-1/2}(1-\alpha^2),
\label{b3bis}
\eeq
where
\beq
\omega\equiv \frac{n\sigma^2g_{11}}{n_\bb\sigma_{12}^2g_{12}}
\label{21}
\eeq
represents the mean free path associated with  cross collisions, relative to the one associated with self collisions. 
The relevant dimensionless control parameters of the problem are the coefficient of restitution $\alpha$, the mass ratio $m/m_\bb$ (or, equivalently, $\mu$), and the dimensionless parameter $\omega$, the latter encompassing the dependence on the densities and the diameters of the granular and  bath particles. This is the parameter space considered in Ref.~\cite{BMP02}. Alternatively, one can eliminate $\omega$ in favor of $\beta$ and use $\alpha$, $m/m_\bb$ (or $\mu$), and $\beta$ as control parameters. This will be the point of view adopted in this paper.

The MB approximation for $T/T_\bb$ consists of inserting Eq.\ (\ref{b2}) into the exact condition (\ref{8}). This leads to a cubic equation for $x$,
\beq
2x(x^2-1)=\beta, 
\label{b3}
\eeq
whose physical root is
\begin{widetext}
\beq
x=\left\{
\begin{array}{ll}
\frac{\sqrt{3}}{3}\left\{\sqrt{3}\cos\left[\frac{1}{3}\sin^{-1}\left(\frac{3\sqrt{3}}{4}\beta\right)\right]+\sin\left[\frac{1}{3}\sin^{-1}\left(\frac{3\sqrt{3}}{4}\beta\right)\right]\right\},&\beta\leq \frac{4\sqrt{3}}{9},\\
\frac{2\sqrt{3}}{3}\cosh\left[\frac{1}{3}\cosh^{-1}\left(\frac{3\sqrt{3}}{4}\beta\right)\right],&\beta\geq \frac{4\sqrt{3}}{9}.
\end{array}
\right.
\label{b4}
\eeq
\end{widetext}
The temperature ratio is then given, on account of Eq.\ (\ref{36}),  by
\beq
\frac{T}{T_\bb}=\frac{m/m_\bb}{(1+m/m_\bb)x^2-1}.
\label{b5}
\eeq
Thus the departure of $x$ from 1 is a measure of the breakdown of the equipartition of energy. It is interesting to note that in this MB approximation the parameter $x$ depends only on $\beta\propto\omega\mu^{-1/2}(1-\alpha^2)$, being independent of the two other parameters ($\alpha$ and $m/m_\bb$). As will be seen in Sec.~\ref{sec3}, this is not so in the Sonine approximation.

Although Eq.\ (\ref{b4}) gives an analytical expression for the temperature ratio in the MB approximation, it is instructive to consider some limiting cases where Eq.\ (\ref{b4}) simplifies considerably.
Let us consider first the limit where the cooling rate is much smaller than the friction constant, i.e.\ $\beta\to 0$. This limit includes as particular cases the quasi-elastic limit ($\alpha\to 1^-$),  the limit of a vanishing mole fraction of granular particles ($n/n_\bb\to 0\Rightarrow\omega\to 0$), and the limit of small granular particles ($\sigma/\sigma_\bb\to 0\Rightarrow\omega\to 0$). In either case, Eq.\ (\ref{b3}) yields 
\beq
x\approx 1+\frac{1}{4} \beta-\frac{3}{32}\beta^2,\quad \beta\to 0 ,
\label{b7bis}
\eeq
and Eq.\ (\ref{b5}) becomes
\begin{subequations}
\beqa
\frac{T}{T_\bb}&\approx& \left[1+\frac{1}{2}(m_\bb/m+1)\beta\right]^{-1},\quad \beta\to 0,
\label{b7a}\\
 &\approx&1-\frac{1}{2}(m_\bb/m+1)\beta,\quad \beta\to 0.
\label{b7b}
\eeqa
\label{b7}
\end{subequations}
Equation (\ref{b7a}) is more general than (\ref{b7b}), since the latter requires that $m/m_\bb$ is kept fixed, in which case the granular temperature is slightly below the bath temperature. However, if in addition to $\beta\to 0$ we have $m/m_\bb\to 0$, Eq.\ (\ref{b7a}) still holds, but (\ref{b7b}) only applies if the product $(m_\bb/m)\beta$ goes to 0. On the other hand, if $\sigma/\sigma_\bb\to 0$ and $m/m_\bb\propto (\sigma/\sigma_\bb)^3$, then $(m_\bb/m)\beta\propto \sigma_\bb/\sigma\to \infty$ and Eq.\ (\ref{b7a}) yields
\beq
T/T_\bb\approx 2(m/m_\bb)\beta^{-1}\propto \sigma/\sigma_\bb, \quad \sigma/\sigma_\bb\to 0.
\label{c4}
\eeq

In the opposite limit of a  cooling rate much larger than the friction constant ($\beta\to \infty$), one gets
\beq
x\approx \left(\frac{1}{2} \beta\right)^{1/3},\quad \beta\to \infty
\label{b8bis}
\eeq
and
\beq
\frac{T}{T_\bb}\approx \frac{m/m_\bb}{1+m/m_\bb}\left(\frac{1}{2} \beta\right)^{-2/3},\quad \beta\to \infty.
\label{b8}
\eeq

A physically important hybrid limit corresponds to $m/m_\bb\to\infty$, $\sigma/\sigma_\bb\to \infty$, and $n/n_\bb\to 0$ in the scaled way $m/m_\bb\propto (\sigma/\sigma_\bb)^3\propto n_\bb/n$, so that the mass densities of the individual granular and bath particles, along with their volume fractions are held fixed. This case was referred to as the ``colloidal'' limit in 
Ref.~\cite{BMP02}. In this scaled limit $\omega\propto \mu\approx m_\bb/m$ and $\beta\propto \mu^{1/2}\to 0$, so that Eqs.\ (\ref{b7bis}) and (\ref{b7b})  apply again.

As will be seen later, the MB approximation provides values for the temperature ratio in good agreement with those obtained from the iterative numerical solution of the Enskog--Boltzmann equation \cite{BMP02}. On the other hand, this approximation is unable to estimate the deviation of the distribution function from the MB form, as measured by the fourth cumulant (or kurtosis) $\kappa$. This requires the use of a Sonine approximation, as done in the next Section.

\section{ Sonine approximation\label{sec3}}
\subsection{Description of the approximation}
The  Sonine approximation \cite{vNE98,GD99} worked out in this Section consists of expanding $f(\mathbf{v})/f_{\text{MB}}(\mathbf{v})$ in Sonine polynomials and then truncating the series after the first non-zero term: 
\beq
f(\mathbf{v})\approx f_{\text{MB}}(\mathbf{v})\left\{1+\frac{\kappa}{2}\left[\frac{15}{4}
-5\frac{mv^2}{2T}+
\left(\frac{mv^2}{2T}\right)^2\right]\right\}.
\label{7}
\eeq
This Sonine approximation is expected to be reliable as long as $|\kappa|$ remains relatively small (say $|\kappa|\lesssim 0.1$). When the approximation (\ref{7}) is used in Eqs.\ (\ref{10}) and (\ref{11}), $A_{12}$ and 
$B_{12}$ become 
 linear functions of $\kappa$, while $A_{11}$ and $B_{11}$ are 
quadratic functions of $\kappa$. However, their quadratic terms are further neglected, in consistency with the spirit of the Sonine approximation (\ref{7}). 
Therefore, 
\begin{subequations}
\beq
A_{ij}\approx A_{ij}^{(0)}+A_{ij}^{(1)}\kappa,
\label{12}
\eeq
\beq
B_{ij}\approx B_{ij}^{(0)}+B_{ij}^{(1)}\kappa.
\label{13}
\eeq
\label{12+13}
\end{subequations}
Inserting (\ref{12+13}) into Eqs.\ (\ref{8}) and (\ref{9}), and 
after eliminating $\kappa$, one gets the following closed equation for the  temperature ratio $T/T_\bb$:
\beq
\left[A_{11}^{(0)}+A_{12}^{(0)}\right]\left[B_{11}^{(1)}+B_{12}^{(1)}\right]=
\left[B_{11}^{(0)}+B_{12}^{(0)}\right]\left[A_{11}^{(1)}+A_{12}^{(1)}\right].
\label{13.1}
\eeq
Once $T$ is determined, the cumulant $\kappa$ is simply given by
\beq
\kappa=-\frac{A_{11}^{(0)}+A_{12}^{(0)}}{A_{11}^{(1)}+A_{12}^{(1)}}=-
\frac{B_{11}^{(0)}+B_{12}^{(0)}}{B_{11}^{(1)}+B_{12}^{(1)}}.
\label{13.2}
\eeq
Equation (\ref{13.2}) provides an \textit{estimate} of the cumulant $\kappa$ within the  first Sonine approximation.
Obviously, the smaller the value of the magnitude of the estimate (\ref{13.2}) the more reliable it is expected to be.

The  expressions for $A_{11}^{(0,1)}$ and $B_{11}^{(0,1)}$ have been derived 
by van Noije and Ernst \cite{vNE98}:
\beq
A_{11}^{(1)}=\frac{3}{16}A_{11}^{(0)},
\label{15}
\eeq
\beq
B_{11}^{(0)}=\frac{2T}{m}\left(\frac{9}{2}+\alpha^2\right)A_{11}^{(0)},
\label{16}
\eeq
\beq
B_{11}^{(1)}=\frac{2T}{m}\left[\frac{3}{32}(69+10\alpha^2)+
\frac{2}{1-\alpha}\right]A_{11}^{(0)},
\label{17}
\eeq
where $A_{11}^\zero$ is given by Eq.\ (\ref{14}).
 The quantities 
$A_{12}^{(0,1)}$ and $B_{12}^{(0,1)}$ were evaluated by Garz\'o and Dufty in Appendices A and B of Ref.\ \cite{GD99}
from the  Sonine approximation in the general case of 
arbitrary coefficients of normal restitution $\alpha_{11}$ and 
$\alpha_{12}$. Particularizing to $\alpha_{12}=1$ we get Eq.\ (\ref{32}) and
\beq
A_{12}^{(1)}=\frac{3}{8}\gamma\frac{2T}{m}
x^{-3}\mu\left[x^2(4-3\mu)-\mu\right],
\label{33}
\eeq
\beq
B_{12}^{(0)}=3\gamma\left(\frac{2T}{m}\right)^2
x^{-1}(1-x^2)
\left[8\mu 
x^4+x^2(5-8\mu)+\mu\right],
\label{34}
\eeq
\beqa
B_{12}^{(1)}&=&\frac{3}{8}\gamma\left(\frac{2T}{m}\right)^2
x^{-5}\left[4x^6(10-33\mu
+48\mu^2-30\mu^3)
\right.
\nonumber\\
&&\left.+
\mu 
x^4(64-87\mu+48\mu^2)
+x^2\mu^2(9\mu-17)+3\mu^3\right].\nonumber\\
&&
\label{35}
\eeqa
Given the values of the three independent parameters $\alpha$, 
 $\mu=(1+m/m_\bb)^{-1}$, and $\beta$ (or, alternatively, $\alpha$, $\mu$, and $\omega$), Eq.\ (\ref{13.1}) yields an algebraic equation of 
tenth-degree for the unknown $x$.
The high degree of this equation  is due to the highly nonlinear dependence of the cross collisional integrals $A_{12}$ and $B_{12}$ on the temperature parameter $x$, as shown by Eqs.\ (\ref{32}) and (\ref{33})--(\ref{35}).
The algebraic  equation (\ref{13.1}) has generally two or four negative real roots (which are unphysical) and two positive real roots, both larger than 1. The largest positive root diverges in the elastic limit ($\alpha\to 1$), while the smallest real root tends to 1 in that limit. Therefore, the physical value of $x$ is  the smallest positive root of Eq.\ (\ref{13.1}).
 Once $x$ 
is determined,  the temperature ratio is given by Eq.\ (\ref{b5}) and
 the cumulant $\kappa$ is obtained from Eq.\ (\ref{13.2}):
\beq
\kappa=\frac{16 x^3[2x(x^2-1)-\beta]}{3\beta x^3+4\mu(4-3\mu)x^2-4\mu^2}.
\label{13.2bis}
\eeq

The physical solution to the tenth-degree equation (\ref{13.1}) is generally very close to the MB approximation (\ref{b4}). In particular, once $\omega$ is eliminated in favor of the parameter $\beta$ defined in Eq.\ (\ref{b3bis}),  the remaining dependence of $x$ on $\alpha$ and $\mu$ 
is generally rather weak. 
On the other hand, it is precisely the difference between the MB and Sonine estimates of $T/T_\bb$, no matter how small it is, what is responsible for the prediction of a non-zero $\kappa$. This is obvious from  Eq.\ (\ref{13.2bis}) and the MB approximation (\ref{b3}).

An approximate solution to Eq.\ (\ref{13.1}) can be obtained by writing 
\beq
x\approx x_0(1+\epsilon),
\label{new1}
\eeq
where $x_0(\beta)$ is the MB solution (\ref{b4}), and then neglecting terms nonlinear in $\epsilon$. The resulting expression for $\epsilon$ is given in the Appendix. In consistency with the assumption (\ref{new1}), Eq.\ (\ref{13.2bis}) becomes
\beq
\kappa\approx \frac{32 x_0^4(3x_0^2-1)}{3\beta x_0^3+4\mu(4-3\mu)x_0^2-4\mu^2}\epsilon .
\label{new2}
\eeq
Equations (\ref{new1}) and \ref{new2}), with $x_0$ given by Eq.\ (\ref{b4}) and $\epsilon$ given by Eqs.\ (\ref{A1})--(\ref{A3}), provide \textit{explicit} expressions for the temperature parameter $x$ and the kurtosis $\kappa$ as functions of $\alpha$, $\mu=(1+m/m_\bb)^{-1}$, and $\beta$. They are useful for practical purposes in those conditions where the MB approximation provides a good estimate of the temperature ratio and so $\kappa$ is relatively small. These include the physically relevant cases of $m/m_\bb \geq 1$ and $\omega\leq 1$ considered in Ref.\ \cite{BMP02}. When $\kappa$ is not small, the numerical solution of Eq.\ (\ref{13.1}) can differ significantly from Eq.\ (\ref{new1}) and so the Sonine approximation may not be applicable any more. This point will be addressed more extensively in Sec.\ \ref{secnew}.
For the sake of completeness, the results presented in the remainder of this paper correspond to the solution of the full Eq.\ (\ref{13.1}), unless  stated otherwise.

\subsection{Limiting behaviors}

In the limit $\beta\to 0$ (i.e.\ $\alpha\to 1^-$ or $\omega\to 0$), Eq.\ (\ref{13.1}) yields the asymptotic behavior 
\beq
x\approx 1+ h \beta,\quad \beta\to 0, 
\label{b7bis2}
\eeq
where
\beq
h=\frac{1}{16}\frac{20-(23+2\alpha^2)\mu+30\mu^2}{5-6\mu+7\mu^2}.
\label{c1}
\eeq
Although the coefficient $h$ is not a constant, it is very close to the numerical value $h\to\frac{1}{4}$ corresponding to the MB approximation (\ref{b7bis}).
The temperature ratio, according to  Eq.\ (\ref{b5}), is given by
\begin{subequations}
\beqa
\frac{T}{T_\bb}&\approx& \left[1+2h(m_\bb/m+1)\beta\right]^{-1},\quad \beta\to 0,
\label{b7a2}\\
 &\approx&1-2h(m_\bb/m+1)\beta,\quad \beta\to 0,
\label{b7b2}
\eeqa
\label{b72}
\end{subequations}
where in the last step it has been assumed that $m_\bb/m$ is finite. 
According to Eq.\ (\ref{13.2bis}), the cumulant $\kappa$ is in this limit
\beq
\kappa\approx \frac{4h-1}{\mu(1-\mu)}\beta, \quad \beta\to 0,
\label{c1bis}
\eeq
provided that the mass ratio is kept fixed. Equations (\ref{b7bis2}) and (\ref{c1bis}) can be obtained as well from Eqs.\ (\ref{new1}) and (\ref{new2}), respectively, by making use of Eqs.\ (\ref{b7bis}) and (\ref{A4}).

Equation (\ref{c1bis}) cannot be applied in the colloidal limit $\beta\propto\mu^{1/2}\to 0$. In that case, it is necessary to evaluate $x$ to second order in $\beta$. The result is
\begin{subequations}
\beq
x\approx 1+\frac{1}{4}\beta-\frac{3(39+2\alpha^2)}{1280}\beta^2,
\label{d3a}
\eeq
\beq
 \frac{T}{T_\bb}
\approx 1-\frac{1}{2}\beta+\frac{3(79+2\alpha^2)}{640}\beta^2.
\label{d3b}
\eeq
\label{d3}
\end{subequations}
The leading term in $x$ coincides with that of the MB estimate (\ref{b7bis}). On the other hand, the subleading term differs from $-\frac{3}{32}$ and is needed to evaluate $\kappa$ from Eq.\ (\ref{13.2bis}) with the result
\beq
\kappa\approx \frac{1-2\alpha^2}{20}\beta.
\label{d4}
\eeq
Again, Eqs.\ (\ref{d3a}) and (\ref{d4}) can also be derived from Eqs.\ (\ref{new1}) and (\ref{new2}), complemented with Eqs.\ (\ref{b7bis}) and (\ref{A5}).
More in general, it can be easily checked from Eqs.\ (\ref{new2}) and (\ref{A7}) that Eq.\ (\ref{d4}) holds when $\beta\to 0$ and $\mu\to 0$, regardless of the scaling law between $\beta$ and $\mu$.
 It is interesting to note that, according to Eq.\ (\ref{d4}), the velocity distribution is platykurtic for $\alpha>\sqrt{2}/2\simeq 0.71$, while it is leptokurtic for $\alpha<\sqrt{2}/2$. The same happens in the case of the homogeneous cooling state, as well as in the case of the white noise thermostat \cite{vNE98}.

In the combined limit $\sigma/\sigma_\bb\to 0$ with $m/m_\bb\propto (\sigma/\sigma_\bb)^3$, i.e. the granular particles are much smaller than the bath particles but the mass densities of the individual particles are comparable in both species, [or more generally if $\beta\to 0$ and $m_\bb/m\to \infty$ with $(m_\bb/m)\beta\to \infty$], Eq.\ (\ref{b7a2}) yields
\beq
T/T_\bb\approx (1/2h)(m/m_\bb)\beta^{-1}\propto\sigma/\sigma_\bb\quad \sigma/\sigma_\bb\to 0,
\label{c42}
\eeq
where in that case $\mu\to 1$ and so $h=(27-2\alpha^2)/96$. In this limit,  Eq.\ (\ref{c1bis}) cannot be used and the cumulant is instead given by
\beq
\kappa\approx \frac{16(4h-1)}{5+2(4h-1)}
=\frac{8(3-2\alpha^2)}{63-2\alpha^2},\quad \sigma/\sigma_\bb\to 0.
\label{c5}
\eeq
This value of $\kappa$ cannot be considered very small ($\frac{8}{61}\leq \kappa\leq \frac{24}{63}$), so the  Sonine approximation, while possibly still qualitatively correct, may not be quantitatively accurate in this limit.
It is worth noting that Eq.\ (\ref{new2}) yields $\kappa\approx   \frac{16}{5}(4h-1)$ in this limit $\mu\to 1$, $\beta\to 0$.

In the (non-colloidal) limit of  heavy granular particles at constant $\omega$ (so $n\sim n_\bb$, $\sigma\sim\sigma_\bb$) we have $m/m_\bb\to \infty\Rightarrow \mu\to 0\Rightarrow \beta\to \infty$. In that case, the solution to Eq.\ (\ref{13.1}) behaves asymptotically as 
\beq
x\approx (\lambda \beta)^{1/3},\quad \beta\to \infty,
\label{b8bis2}
\eeq
where the amplitude $\lambda$ is
\beq
\lambda={2}\frac{16+(45+6\alpha^2)(1-\alpha)}{64+(177+30\alpha^2)(1-\alpha)}.
\label{c2}
\eeq
This parameter is very close to the numerical value $\lambda\to \frac{1}{2}$ given by the MB approximation (\ref{b8bis}).
In this limit the temperature ratio $T/T_\bb$ goes to zero as
\beq
\frac{T}{T_\bb}\approx (\lambda \beta)^{-2/3},\quad \beta\to \infty,
\label{b82}
\eeq
and the cumulant tends to the (small) constant value
\beqa
\kappa&\approx& \frac{16}{3}(2\lambda-1)\nonumber\\
&=&\frac{16(1-\alpha)(1-2\alpha^2)}{64+(177+30\alpha^2)(1-\alpha)}
,\quad m/m_\bb\to\infty.\nonumber\\
&&
\label{c3}
\eeqa
Interestingly, this expression coincides with the one derived by van Noije and Ernst for the white noise thermostat \cite{vNE98}. 
For this reason, I will call this limit the ``white noise'' limit.
I will come back to this point in Sec.\ \ref{sec4}.
In this white noise limit, Eq.\ (\ref{new2}) becomes $\kappa\approx \epsilon/16$, with $\epsilon$ given by Eq.\ (\ref{A8}).

The expressions for $T/T_\bb$ and $\kappa$ in the limits discussed above are summarized in Table \ref{table1}.
\begin{table*}
\caption{\label{table1} Summary of the expressions for the temperature ratio $T/T_\bb$ and the kurtosis $ \kappa$ in the Sonine approximation for some limiting situations where $\beta$ tends to 0 or to $\infty$. The parameter $\beta$ is a measure of the cooling rate $\zeta$ relative to the friction constant $\gamma$ and is a function of the coefficient of restitution $\alpha$, the size ratio $\sigma/\sigma_\bb$, the number densities $n$ and $n_\bb$, and the mass ratio $m/m_\bb$ [cf. Eqs.\ (\protect\ref{b3bis}) and (\protect\ref{21})]. The coefficients $h$ and $\lambda$ are given by Eqs.\ (\protect\ref{c1}) and (\protect\ref{c2}), respectively. Case 1 represents the quasi-elastic limit. In cases 2 and 3 the mean free path associated with cross collisions is much smaller than the one associated with self collisions ($\omega\to 0$), but the masses of the granular and bath particles are comparable. In case 4 the granular particles are much smaller and much lighter than the bath particles. In case 5 the granular particles are much larger and much heavier than the bath particles but both species occupy a comparable volume fraction (colloidal limit). Finally, in case 6 the granular particles are much heavier than the bath particles but the mean free path ratio $\omega$ is finite (white noise limit).}
\begin{ruledtabular}
\begin{tabular}{cccccccc}
Case&$\alpha$&$\sigma/\sigma_\bb$&$n/n_\bb$&$m/m_\bb$&$\beta$&$T/T_\bb$&$\kappa$\\
\hline
1&$\to 1^-$&fixed&fixed&fixed&$\sim(1-\alpha)\to 0$&$1-2h(m_\bb/m+1)\beta$&$(4h-1)\frac{(1+m/m_\bb)^2}{m/m_\bb}\beta$\\
2&fixed&fixed&$\to 0$&fixed&$\sim n/n_\bb\to 0$&$1-2h(m_\bb/m+1)\beta$&$(4h-1)\frac{(1+m/m_\bb)^2}{m/m_\bb}\beta$\\
3&fixed&$\to 0$&fixed&fixed&$\sim(\sigma/\sigma_\bb)^2\to 0$&$1-2h(m_\bb/m+1)\beta$&$(4h-1)\frac{(1+m/m_\bb)^2}{m/m_\bb}\beta$\\
4&fixed&$\to 0$&fixed&$\sim (\sigma/\sigma_\bb)^3\to 0$&$\sim (\sigma/\sigma_\bb)^2\to 0$&$\frac{48}{27-2\alpha^2}(m/m_\bb)\beta^{-1}\to 0$&$8\frac{3-2\alpha^2}{63-2\alpha^2}$\\
5&fixed&$\to\infty$&$\sim (\sigma/\sigma_\bb)^{-3}\to 0$&$\sim (\sigma/\sigma_\bb)^3\to \infty$&$\sim (\sigma/\sigma_\bb)^{-3/2}\to 0$&$1-\frac{1}{2}\beta$&$\frac{1-2\alpha^2}{20}\beta$\\
6&fixed&fixed&fixed&$\to\infty$&$\sim(m/m_\bb)^{1/2}\to\infty$&$(\lambda\beta)^{-2/3}\to 0$&$\frac{16}{3}(2\lambda-1)$
\end{tabular}
\end{ruledtabular}
\end{table*}

\subsection{Transient regime}
 \begin{figure}
 \includegraphics[width=.90 \columnwidth]{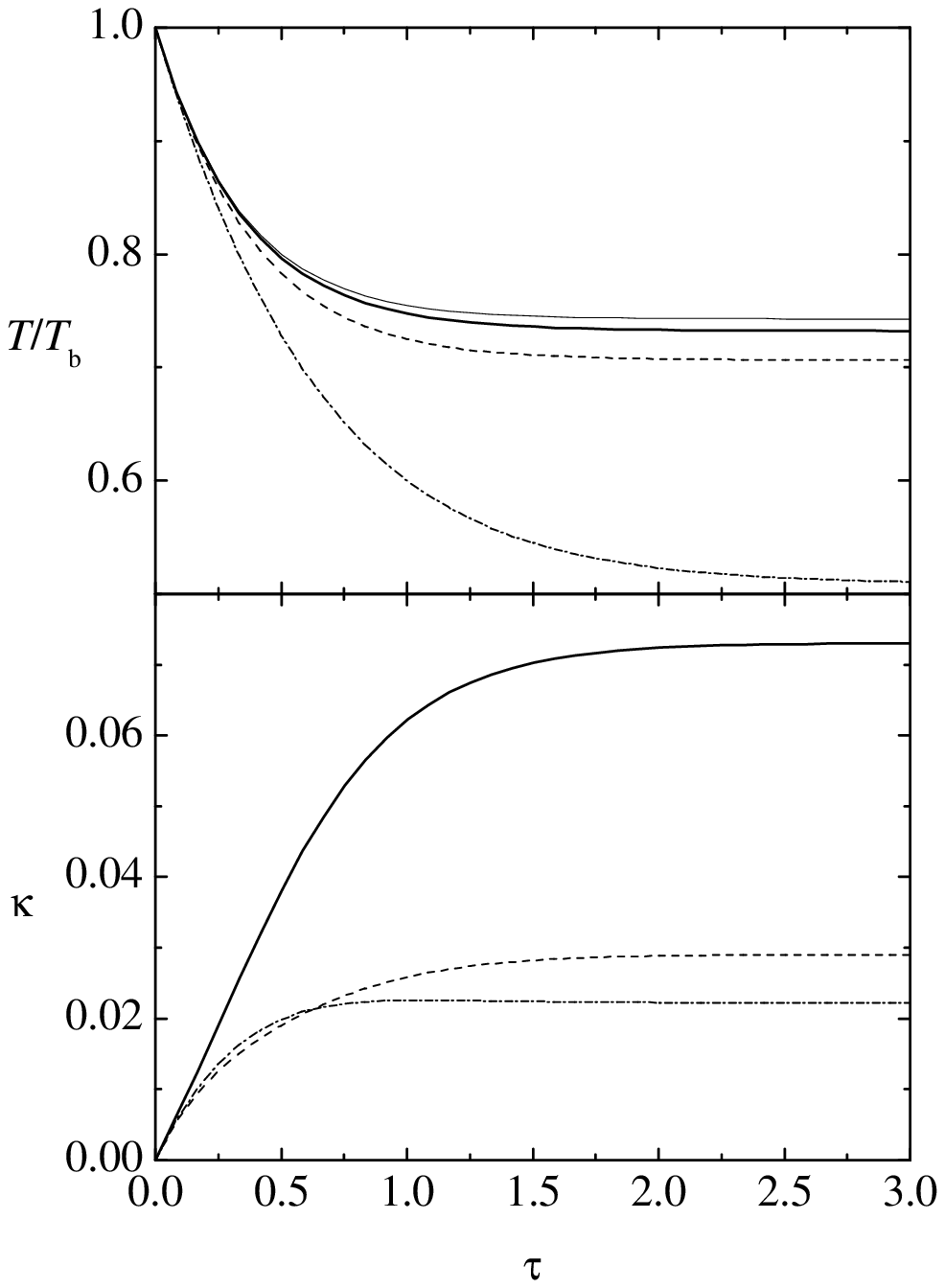}
 \caption{Time evolution of the temperature ratio $T/T_\bb$ and the cumulant $\kappa$, according to the  Sonine approximation, for $\alpha=0.5$, $\omega=1$, and $m/m_\bb=1$ (thick solid line), 10 (dash line), and 100 (dash-dot line). The initial condition is $T/T_\bb=1$ and $\kappa=0$. The thin solid line in the upper panel represents the time evolution of the temperature ratio in the MB approximation for the case $m/m_\bb=1$. For the other two cases the MB curves are indistinguishable from the Sonine ones.\label{evol}}
 \end{figure}
The  Sonine approximation (\ref{7}) not only allows one  to get the steady-state values of the temperature ratio $T/T_\bb$ and the cumulant $\kappa$, but also their time evolution. Let us introduce the scaled time variable
\beq
\tau(t)=\int_0^t\dd t' \nu(t'),
\label{c7}
\eeq
which counts time essentially as the average number of self collisions experienced by a granular particle. In that case, Eqs.\ (\ref{8bis}) and (\ref{9bis}) can be rewritten as
\begin{subequations}
\beq
\partial_\tau T=-\frac{2}{3}A^* T,
\label{c8}
\eeq
\beq
\partial_\tau \kappa=-\frac{4}{15}B^*+\frac{4}{3}(\kappa+1)A^*,
\label{c9}
\eeq
\label{c8+c9}
\end{subequations}
where
\beq
A^*\equiv \frac{A_{11}+A_{12}}{\nu(2T/m)}, \quad B^*\equiv \frac{B_{11}+B_{12}}{\nu(2T/m)^2}.
\label{c10}
\eeq
Equations (\ref{c8+c9}) are exact, but they do not form a closed set. On the other hand, if the approximation (\ref{12+13}) is used, Eqs.\ (\ref{c8+c9}) become a set of two nonlinear autonomous equations that can be solved numerically by specifying the initial values of $T/T_\bb$ and $\kappa$ only.
As an illustration, Fig.~\ref{evol} shows the time evolution of $T/T_\bb$ and $\kappa$ for $\alpha=0.5$, $\omega=1$, and $m/m_\bb=1$, 10, and 100, starting from an initial condition with $T/T_\bb=1$ and $\kappa=0$. We observe that after a few  collisions per particle the steady state has been reached. Figure \ref{evol} also shows the time evolution of the temperature ratio in the MB approximation for the case $m/m_\bb=1$. For the other two cases the MB curves are indistinguishable from the Sonine ones, in consistency with the fact that $\kappa$ is rather small in those two cases.
\section{Comparison with the numerical solution of Biben et al.\ \protect\cite{BMP02}\label{sec5}}
\begin{figure}[t]
 \includegraphics[width=.90 \columnwidth]{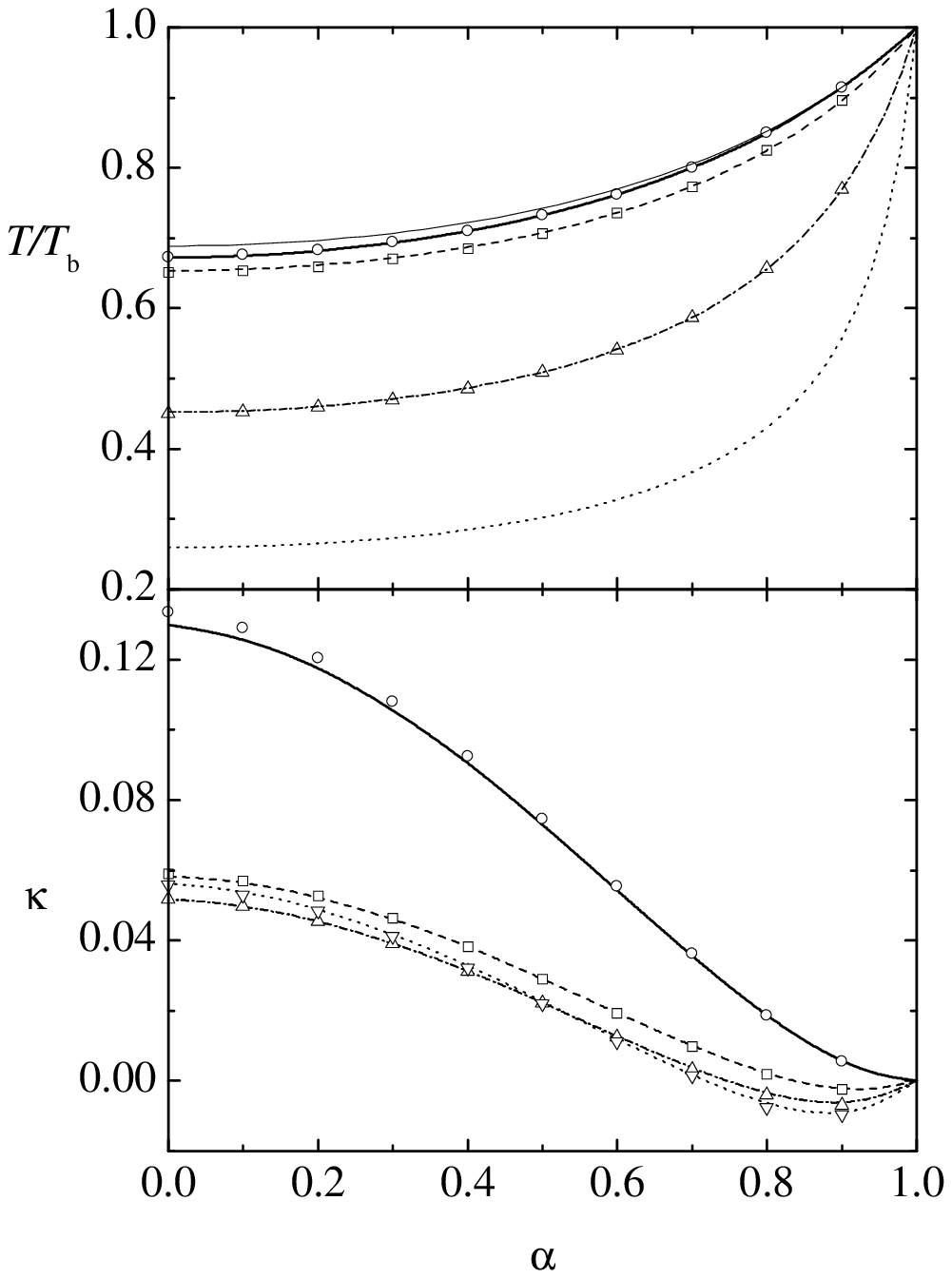}
 \caption{Temperature ratio $T/T_\bb$ and cumulant $\kappa$ versus $\alpha$ for $\omega=1$ and $m/m_\bb=1$ (thick solid lines and circles), 10 (dash lines and squares), 100 (dash-dot lines and up triangles), and 1000 (dot lines and down triangles). The lines correspond to the  Sonine approximation, while the symbols correspond to the numerical solution of the Enskog--Boltzmann equation \protect\cite{BMP02}. (The values of the temperature ratio for $m/m_\bb=1000$  are not given in Ref.\ \protect\cite{BMP02}.)
The thin solid line in the top panel represents the  temperature ratio in the MB approximation for the case $m/m_\bb=1$. For the other two cases the MB curves are indistinguishable from the Sonine ones and are not plotted. \label{noise}}
 \end{figure}

By an iterative numerical scheme, Biben et al.\ have solved the (steady-state) Enskog--Boltzmann equation (\ref{2}) for several choices of the mean free path ratio $\omega\leq 1$, the mass ratio $m/m_\bb\geq 1$, and the coefficient of restitution $\alpha< 1$. 
Figure \ref{noise} compares $T/T_\bb$ and $\kappa$ in the  Sonine approximation with the numerical solution for $\omega=1$ and $m/m_\bb=1$, 10, 100, and 1000. As can be seen, the agreement is excellent, except for a small tendency of the Sonine approximation to underestimate $\kappa$ for $m/m_\bb=1$ and large inelasticity. For the other three values of the mass ratio, the cumulant $\kappa$ is small enough to make the temperature ratio predicted by the MB approximation practically indistinguishable from that predicted by the Sonine approximation. The sequence of increasing values of $m/m_\bb$  at fixed $\omega$ in Fig.\ \ref{noise} leads to the white noise limit discussed in the previous Section. In fact, the asymptotic expression (\ref{c3}) describes very accurately the curve of $\kappa$ for $m/m_\bb =1000$.

\begin{figure}[t]
 \includegraphics[width=.90 \columnwidth]{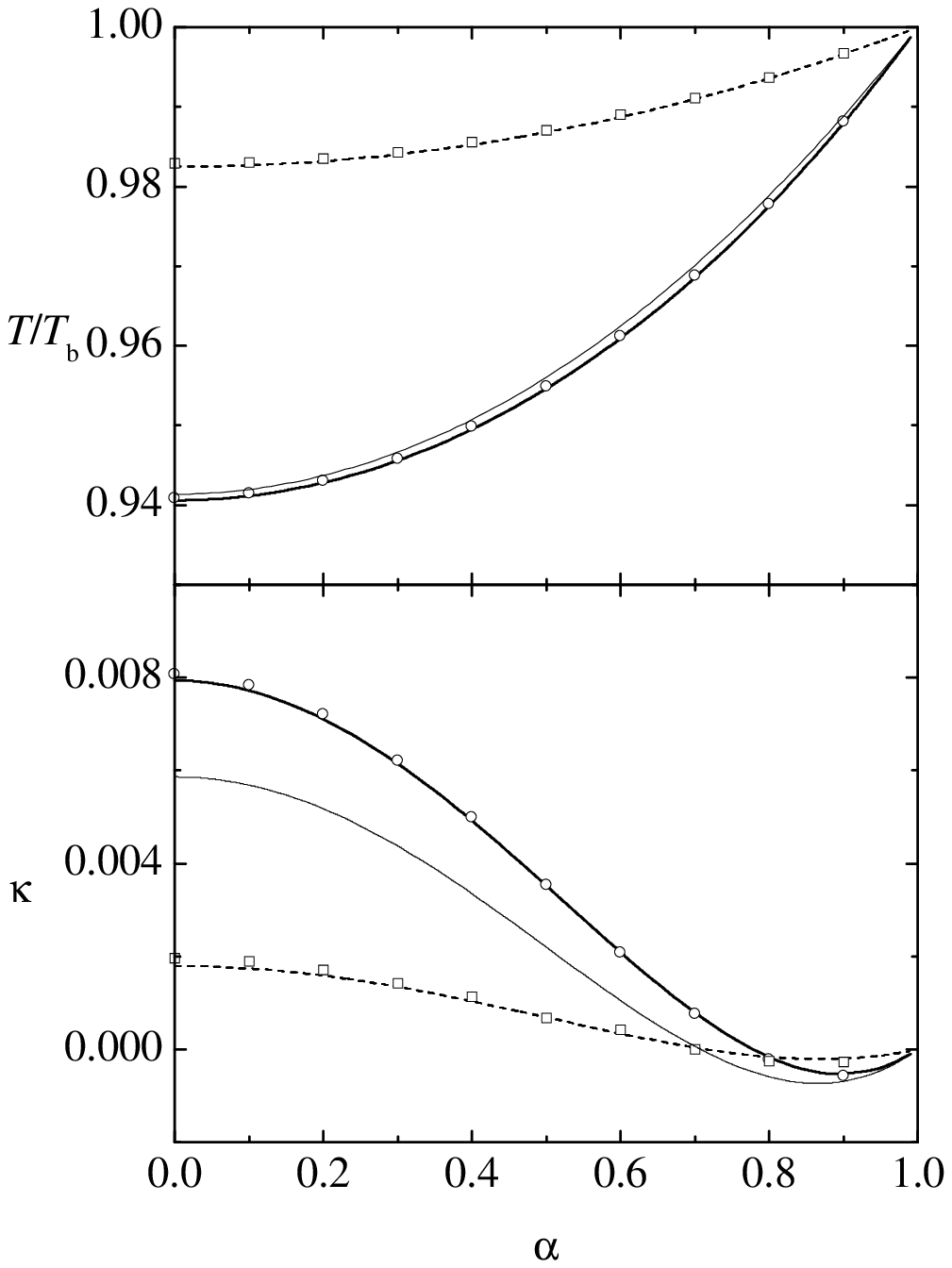}
 \caption{Temperature ratio $T/T_\bb$ and cumulant $\kappa$ versus $\alpha$ for $(m/m_\bb)\omega=1$ with $m/m_\bb=10$ (thick solid lines and circles) and 100 (dash lines and squares). The lines correspond to the  Sonine approximation, while the symbols correspond to the numerical solution of the Enskog--Boltzmann equation \protect\cite{BMP02}.
The temperature ratios predicted by the MB approximation are practically indistinguishable from the Sonine ones and are not plotted.
The thin solid lines represent the asymptotic behaviors given by the leading term of Eq.\ (\protect\ref{d3b}) (top panel) and by Eq.\ (\protect\ref{d4}) (bottom panel) for the case $m/m_\bb=10$. For $m/m_\bb=100$ the asymptotic behaviors   are indistinguishable from the Sonine curves and are not plotted.
 \label{colloidal}}
 \end{figure}

A sequence leading to the colloidal limit ($\omega\propto m_\bb/m$, $m/m_\bb \to\infty$) is shown in Fig.~\ref{colloidal}, where $T/T_\bb$ and $\kappa$ are plotted for fixed $(m/m_\bb)\omega=1$ with $m/m_\bb=10$ and 100.
Again, the agreement with the numerical solution is excellent. For $m/m_\bb=10$ and 100, the temperature ratio is already very well described by the asymptotic law $T/T_\bb=1-\frac{1}{2}\beta$ [cf. Eq.\ (\ref{d3b})]. As for $\kappa$, the asymptotic behavior (\ref{d4}) has been practically reached for $m/m_\bb=100$. 

\begin{figure}
 \includegraphics[width=.90 \columnwidth]{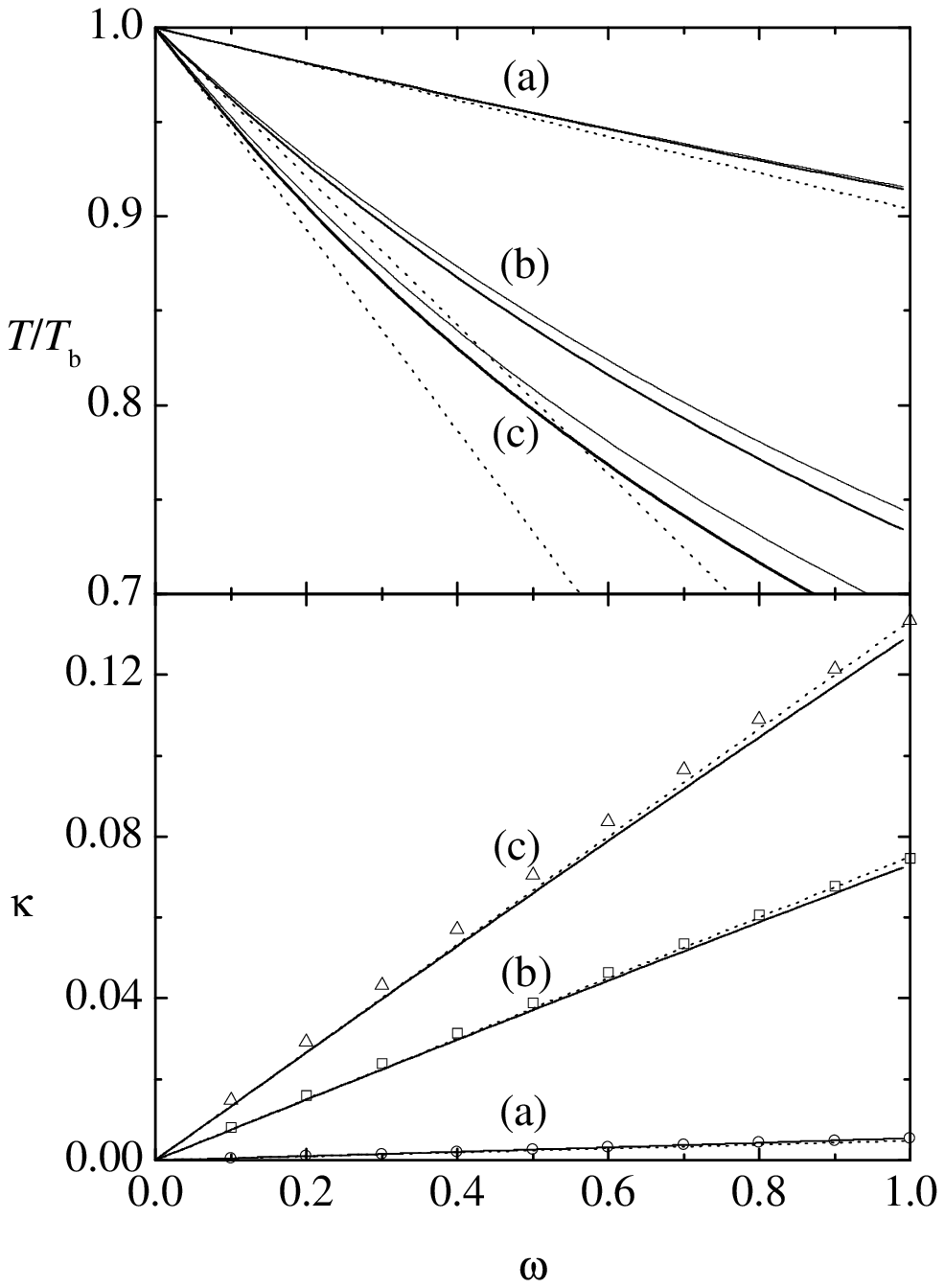}
 \caption{Temperature ratio $T/T_\bb$ and cumulant $\kappa$ versus $\omega$ for $m/m_\bb=1$ and (a) $\alpha=0.9$, (b) $\alpha=0.5$, and (c) $\alpha=0$. The thick solid lines correspond to the  Sonine approximation, while the symbols in the bottom panel correspond to the numerical solution of the Enskog--Boltzmann equation \protect\cite{BMP02}. 
(The values of the temperature ratio for cases (a)--(c)  are not given in Ref.\ \protect\cite{BMP02}.)
The thin solid lines in the top panel represent the  temperature ratio in the MB approximation. The dotted lines in both panels are the asymptotic linear behaviors (\protect\ref{b7b2}) and  (\protect\ref{c1bis}).
 \label{omega}}
 \end{figure}

Let us consider now the dependence of $T/T_\bb$ and $\kappa$ on the mean free path ratio $\omega$ for fixed values of $m/m_\bb$ and $\alpha$. Figure \ref{omega} shows the results for equal masses ($m/m_\bb=1$) and three values of the coefficient of restitution ($\alpha=0.9$, 0.5, and 0). Near the origin, $\beta\propto\omega$ is small and then the asymptotic laws (\protect\ref{b7b2}) and  (\protect\ref{c1bis}) apply. As the inelasticity increases, the range of values of $\omega$ for which the temperature decays linearly shrinks. On the other hand, since,
as remarked by Biben et al.\ \cite{BMP02}, the cumulant is practically a linear function of $\omega$, the asymptotic law (\protect\ref{c1bis}) keeps being rather accurate even near $\omega=1$.

\begin{figure}
 \includegraphics[width=.90 \columnwidth]{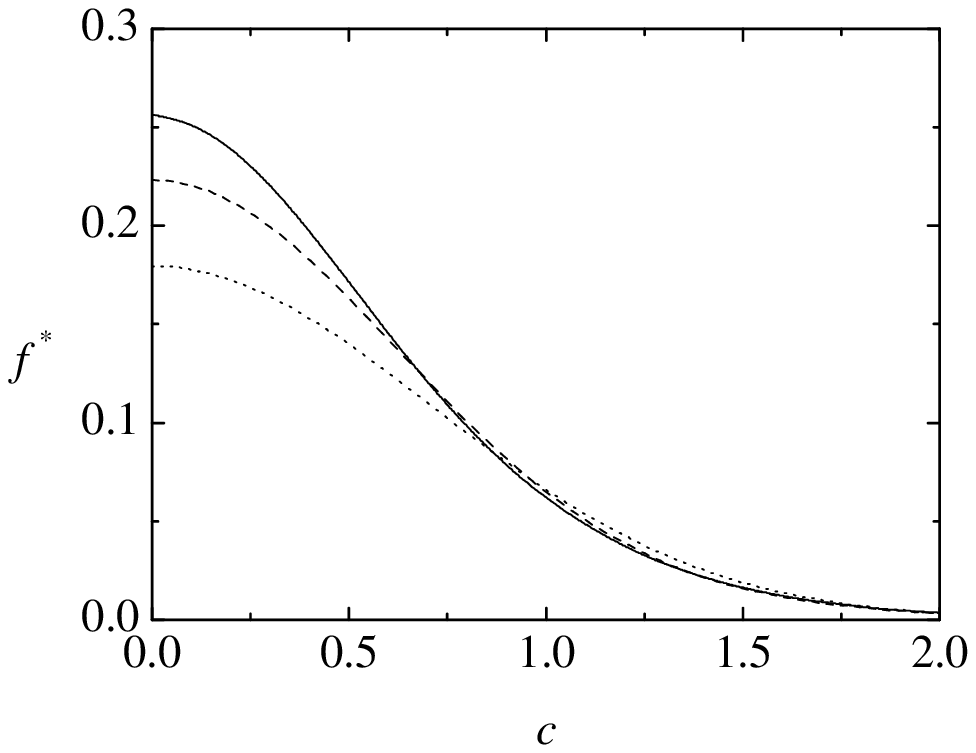}
 \caption{Reduced velocity distribution function $f^*(c)$ for $\omega=1$, $m/m_\bb=1$, and $\alpha=0$. The solid line is the numerical solution of the Enskog--Boltzmann equation \protect\cite{BMP02}, while the dashed line is the  Sonine approximation (\ref{7}) with $\kappa=0.12984$. The dotted line is the MB distribution $f^*(c)=\pi^{-3/2} \e^{-c^2}$.
 \label{distrib}}
 \end{figure}

As is apparent from Figs.\ \ref{noise}--\ref{omega}, the  Sonine approximation yields excellent estimates for the temperature ratio $T/T_\bb$ and the kurtosis $\kappa$ for the cases considered in Ref.~\cite{BMP02}. Even the MB approximation does a good job in estimating $T/T_\bb$, as first noted by Barrat and Trizac \cite{BT02a}. 
Because of this, the Sonine curves plotted in Figs.\ \ref{noise}--\ref{omega}, which have actually been obtained from Eq.\ (\ref{13.2bis}) and the numerical solution to Eq.\ (\ref{13.1}), are practically indistinguishable from those obtained from the approximate solution (\ref{new1}) and (\ref{new2}).
Does the good performance of the Sonine approximation mean that the distribution function is accurately described by the truncated expansion (\ref{7})? Not necessarily so if $\kappa$ is not very small, as illustrated in Fig.~\ref{distrib}, where the reduced distribution function
\beq
f^*(\mathbf{c})=n^{-1}(2T/m)^{3/2}f(\mathbf{v}), \quad \mathbf{c}=\mathbf{v}/\sqrt{2T/m},
\label{e1}
\eeq
is plotted for the case  with the largest deviation from equilibrium considered in Ref.~\cite{BMP02}, namely equal masses ($m/m_\bb=1$), equal mean free paths ($\omega=1$), and totally inelastic collisions ($\alpha=0$). 
Figure \ref{distrib} shows that  the  Sonine approximation (\ref{7}) \textit{partially} accounts for the departure of the true distribution function from the equilibrium one. The remaining difference is due to Sonine polynomials of order higher than two. Notwithstanding this, the truncated expansion (\ref{7}) is still useful for estimating the collisional integrals  (\ref{10}) and (\ref{11}). Of course, for other choices of the control parameters such that $\kappa$ is smaller than in the case considered in Fig.~\ref{distrib}, the approximate distribution (\ref{7}) is expected to be much more accurate.

\section{Fokker--Planck limit. High energy tail\label{sec4}}
\subsection{Limit $m/m_\bb\to\infty$}
A physically important situation arises when the granular particles are much heavier than the bath particles. This includes the colloidal limit [$m/m_\bb\to \infty$, $\omega\propto m_\bb/m\Rightarrow\beta\propto (m_\bb/m)^{1/2}\to 0$], as well as the white noise limit [$m/m_\bb\to \infty$, $\omega=\text{finite}\Rightarrow \beta\propto (m/m_\bb)^{1/2}\to \infty$]. 
The predictions of the  Sonine approximation in these limits have been analyzed in the previous Section. Nonetheless, it is worth studying the limit $m/m_\bb\to \infty$ at the level of the kinetic equation itself.

If $m/m_\bb\to\infty\Rightarrow \mu\approx m_\bb/m\to 0$, the 
Enskog--Boltzmann collision operator $J_{12}[\mathbf{v}|f_1,f_2]$ can be approximated by the 
Fokker--Planck operator (see, for instance, Appendix A of Ref.\ \cite{BDS99}),
\beq
J_{12}[\mathbf{v}|f_1,f_2]\to \gamma\sqrt{\frac{T_\bb}{T}}
\frac{\partial}{\partial\mathbf{v}}\cdot\left(\mathbf{v}+\frac{T_\bb}{m}
\frac{\partial}{\partial\mathbf{v}}\right)f(\mathbf{v}),
\label{18}
\eeq
where $\gamma$ is given by Eq.\ (\ref{19}). This confirms the interpretation of $\gamma$ as a friction constant. 
In the Fokker--Planck limit (\ref{18}) the quantities $A_{12}$ and $B_{12}$ 
can be evaluated \textit{exactly},
\begin{subequations}
\beq
A_{12}=3\gamma \frac{2T}{m}\sqrt{\frac{T_\bb}{T}}\left(1-\frac{T_\bb}{T}\right),
\label{22}
\eeq
\beq
B_{12}=15\gamma\left(\frac{2T}{m}\right)^2\sqrt{\frac{T_\bb}{T}}\left(1-\frac{T_\bb}{T}+\kappa\right).
\label{23}
\eeq
\label{22+23}
\end{subequations}
This means that in this limit the Sonine approximation (\ref{12+13}) for $(i,j)=(1,2)$ becomes exact. In fact, Eqs.\ (\ref{22+23}) can be reobtained from Eqs.\ (\ref{32}) and (\ref{33})--(\ref{35}) by taking the limit $\mu\to 0$, in which case $x\to (T_\bb/T)^{1/2}$.
On the other hand, the self collision operator $J_{11}$ is obviously not affected by the limit $m/m_\bb\to\infty$, so that Eqs.\ (\ref{12+13}) for $(i,j)=(1,1)$ are still approximate.
In the MB approximation, the parameter $x=(T_\bb/T)^{1/2}$ keeps obeying the cubic equation (\ref{b3}). On the other hand, in the Sonine approximation the tenth-degree equation (\ref{13.1}) simplifies to a quartic equation:
\begin{widetext}
\beq
320(1-\alpha)x^2(x^2-1)+\beta\left[64+(177+30\alpha^2)(1-\alpha)\right]
 x(x^2-1)-160\beta(1-\alpha)x-2\beta^2\left[16+(45+6\alpha^2)(1-\alpha)\right]=0.
\label{d1}
\eeq
\end{widetext}
Moreover, Eq.\ (\ref{13.2bis}) reduces to
\beq
\kappa=\frac{16}{3\beta}\left[2x(x^2-1)-\beta\right].
\label{d2}
\eeq
The physical root of Eq.\ (\ref{d1}) can be obtained analytically.
Nevertheless, a simpler practical expression is provided by  Eq.\ (\ref{new1}) with $\epsilon$ given by Eq.\ (\ref{A6}).

The above results apply in the limit $m/m_\bb\to\infty$ regardless of the magnitude of $\beta$. In particular, $\beta$ remains finite if $\omega\to 0$ as $m/m_\bb\to\infty$ in the scaled way $\omega\propto (m_\bb/m)^{1/2}$. This may correspond to $n_\bb/n \propto (\sigma/\sigma_\bb)^3$ and $m/m_\bb\propto (\sigma/\sigma_\bb)^{6}$, so that the volume fractions of both species are comparable but the mass density of a granular particle is much larger than that of a bath particle. Below  the more interesting cases of the colloidal limit and the white noise limit (where $\beta\to 0$ and $\beta\to\infty$, respectively) are considered. 

\subsection{Colloidal limit} 

Let us assume that the heavy granular particles are made of a material with  a mass density comparable to that of the bath particles, i.e.\ $m/m_\bb\propto (\sigma/\sigma_\bb)^3$. If, in addition, the partial volume fractions of both species are comparable, we have $n/n_\bb\propto (\sigma_\bb/\sigma)^3$. Consequently, $\omega\propto m_\bb/m$ and $\beta\propto (m_\bb/m)^{1/2}\to 0$.
In that case, Eqs.\ (\ref{d1}) and (\ref{d2}) reduce to Eqs.\ (\ref{d3a}) and (\ref{d4}), respectively.
Therefore, the same results are obtained in the colloidal limit following two routes: (i) first $m/m_\bb\to \infty$ at fixed $\beta$ and then $\beta\to 0$; (ii) $m/m_\bb\to \infty$ and $\beta\to 0$ simultaneously, with $\beta(m/m_\bb)^{1/2}$ fixed. 

\subsection{White noise limit}

The Fokker--Planck limit (\ref{18}) assumes that $m/m_\bb\to \infty$ but otherwise the 
parameter $\omega$, essentially measuring the concentration ratio $n/n_\bb$, 
remains arbitrary. Now we consider that $\omega$ is finite, so that $\beta\propto (m/m_\bb)^{1/2}\to\infty$.
In this situation 
where the concentration of the heavy granular particles is comparable to 
that of the bath particles, the cooling rate of the granular fluid is much 
larger than the friction constant and, as a consequence, the steady state is 
reached with a granular temperature much smaller than the bath temperature.
In this case, Eqs.\ (\ref{d1}) and (\ref{d2}) yield Eqs.\ (\ref{b8bis2}) and (\ref{c3}), with $\lambda$ given by Eq.\ (\ref{c2}), and the temperature is given by Eq.\ (\ref{b82}).
As said in Sec.~\ref{sec3}, the result (\ref{c3}) coincides exactly with that obtained in the case of a granular gas heated with a white noise thermostat \cite{vNE98}.
To understand this, note that
since $T_\bb/T\gg 1$, the drift term can be neglected versus the diffusion 
term in the Fokker--Planck operator (\ref{18}) (for $v\ll \sqrt{2T_\bb/m}$) with the result 
\beq
J_{12}[\mathbf{v}|f_1,f_2]\to \frac{\xi_0^2}{2}
\left(\frac{\partial}{\partial\mathbf{v}}\right)^2f(\mathbf{v}),
\label{27}
\eeq
where
\beqa
\xi_0^2&=&\frac{2T}{m}\gamma\left({\frac{T_\bb}{T}}\right)^{3/2}\nonumber\\
&=&\frac{2T}{m}\zeta^{(0)}\lambda,
\label{28}
\eeqa
where in the last step use has been made of Eq.\ (\ref{b82}) and the definition $\beta=\zeta^{(0)}/\gamma$.
The right-hand side of (\ref{27}) can be interpreted as representing the 
effect of an external white noise force $\mathbf{F}(t)$ with the properties 
\cite{vNE98,WM96,W96,SBCM98}
\beq
\langle \mathbf{F}(t)\rangle =0,\quad \langle 
\mathbf{F}(t)\mathbf{F}(t')\rangle =m^2\xi_0^2\mathsf{I}\delta(t-t'),
\label{29}
\eeq
where $\mathsf{I}$ is the $3\times 3$ unit matrix.
Now we see that this white noise thermostat can be obtained as a limiting case of a thermostat consisting of a bath of elastic particles.
The stochastic force  $\mathbf{F}(t)$  can be interpreted  as arising from 
collisions of the granular particles with a comparable number of much 
lighter (and hotter) bath particles. The first line of Eq.\ (\ref{28}) relates the strength of the force correlations with  the size, mass, density, and temperature of the bath particles, and with the size and mass of the granular particles [cf. Eq.\ (\ref{19})]. The second line holds in the steady state and relates $\xi_0^2$ with the properties of the granular fluid only.

\subsection{High energy tail}

The Fokker--Planck limit (\ref{18}) allows us to speculate on the high energy tail of the velocity distribution function of the granular particles. Let us assume that
\beq 
f(\mathbf{v})\approx K \e^{-k v^a},\quad v\gg \sqrt{2T/m},
\label{d5}
\eeq
where $K$, $k$, and $a$ are unknown constants. If $a<2$, it can be argued that \cite{vNE98,EP97}
\beq
J_{11}[\mathbf{v}|f,f]\approx -\pi n \sigma^2g_{11}v f,\quad v\gg \sqrt{2T/m}.
\label{d6}
\eeq
Inserting (\ref{d5}) into the Fokker--Planck operator (\ref{18}), we have
\beqa
\frac{\partial}{\partial\mathbf{v}}\cdot\left(\mathbf{v}+\frac{T_\bb}{m}
\frac{\partial}{\partial\mathbf{v}}\right)f(\mathbf{v})&\approx& ak v^a\left(
\frac{T_\bb}{m}ak v^{a-2}-1\right)f,\nonumber\\
&& k v^a\gg 1.
\label{d7}
\eeqa
To fix ideas, let us consider those cases such  that in the steady state $T\ll T_\bb$ (i.e.\ $\beta\gg 1$). This happens, for instance, in the white noise limit. Then, there exists an intermediate regime of velocities with
$\sqrt{2T/m}\ll v\ll \sqrt{2T_\bb/m}$. Let us assume (to be checked by consistency later) that in that case $(T_\bb/m)ak v^{a-2}\gg 1$, so the balance between $J_{11}$ and $J_{12}$ yields $a=\frac{3}{2}$ and
\beq
k^2=\frac{4\pi mn\sigma^2g_{11}}{9T\gamma(T_\bb/T)^{3/2}}=\frac{2\sqrt{2\pi}}{3\lambda(1-\alpha^2)}\left(\frac{m}{2T}\right)^{3/2},
\label{d8}
\eeq
where in the last step use has been made of Eqs.\ (\ref{17.1}), (\ref{17.2}), and (\ref{b82}).
The condition $k v^a\gg 1$ is consistently satisfied if $v\gg \sqrt{2T/m}$ because $k v^a\sim (mv^2/2T)^{3/4}$. As for the consistency of the assumed condition $(T_\bb/m)ak v^{a-2}\gg 1$, we have
\beq
\frac{T_\bb}{m}ak v^{a-2}\sim \left(\frac{mv^2}{2T}\right)^{3/4} \left(\frac{mv^2}{2T_\bb}\right)^{-1},
\label{d9}
\eeq
so this quantity is indeed large in the intermediate regime $(mv^2/2T)\gg 1\gg (mv^2/2T_\bb)$.
In summary, we get
\beq 
f(\mathbf{v})\approx K \e^{-k v^{3/2}},\quad \sqrt{2T/m}\ll v\ll \sqrt{2T_\bb/m},
\label{d10}
\eeq
with $k$ given by Eq.\ (\ref{d8}).

On the other hand, in the true regime of asymptotically large velocities, $v\gg \beta \sqrt{2T_\bb/m}$, the term $J_{11}$ can be neglected versus the diffusion and drift terms of the Fokker--Planck operator $J_{12}$  and therefore $a=2$, $k=m/2T_\bb$. Thus,
\beq 
f(\mathbf{v})\approx K' \e^{-mv^2/2T_\bb },\quad  v\gg \beta \sqrt{2T_\bb/m}\gg \sqrt{2T/m}.
\label{d11}
\eeq
Strictly speaking, the asymptotic behavior (\ref{d6}) does not hold if $a=2$, but it is still plausible that $|J_{11}[\mathbf{v}|f,f]|\ll \gamma \sqrt{T_\bb/T}(mv^2/2T_\bb)f$ if $v\gg \beta \sqrt{2T_\bb/m}$
(i.e.\ the most energetic granular particles are not affected by 
self collisions). 
The crossover from the stretched exponential (\ref{d10}) to the Gaussian decay (\ref{d11}) takes place around a velocity $v_*$ that can be estimated from the condition $mv_*^2/2T_\bb\sim k v_*^{3/2}$, with $k$ given by Eq.\ (\ref{d8}). This yields
\beq
\left(\frac{mv_*^2}{2T}\right)^{1/2}\sim \frac{2\sqrt{2\pi}}{3\lambda(1-\alpha^2)}\left(\frac{T_\bb}{T}\right)^{2}.
\label{d12}
\eeq
This crossover velocity is indeed much larger than the thermal velocity because $T\ll T_\bb$.

The above analysis is not strictly valid if there is no a great disparity between both temperatures, so that the intermediate regime (\ref{d10}) does not exist. 
It is reasonable to conjecture that the high energy tail is of the form (\ref{d5}) with $a=2$ but $k\neq m/2T_\bb$, since in that case Eq.\ (\ref{d6}) does not hold and one may expect that $|J_{11}[\mathbf{v}|f,f]|\sim \nu(mv^2/2T)f$ instead. 
Moreover, if $m\geq m_\bb$ but the masses are not very disparate, the Fokker--Planck limit (\ref{18}) is not valid either. On the other hand, the results obtained here, along with the evidence of the numerical solution of Ref.~\cite{BMP02} (see Fig.~8 of Ref.~\cite{BMP02}) give support to the following scenario: For velocities $v\gg \sqrt{2T_\bb/m}$, the distribution function reaches an asymptotic form (\ref{d5}) with $a=2$, i.e. it is a Gaussian; if, in addition, $T\ll T_\bb$, there exists a window of velocities much larger than the thermal velocity $\sqrt{2T/m}$ but much smaller than $\sqrt{2T_\bb/m}$ where the asymptotic Gaussian form has not been reached yet and the distribution function can be approximated by the stretched exponential (\ref{d10}).

\section{Range of applicability of the Sonine approximation. Critical behavior}
\label{secnew}
\begin{figure}
 \includegraphics[width=.90 \columnwidth]{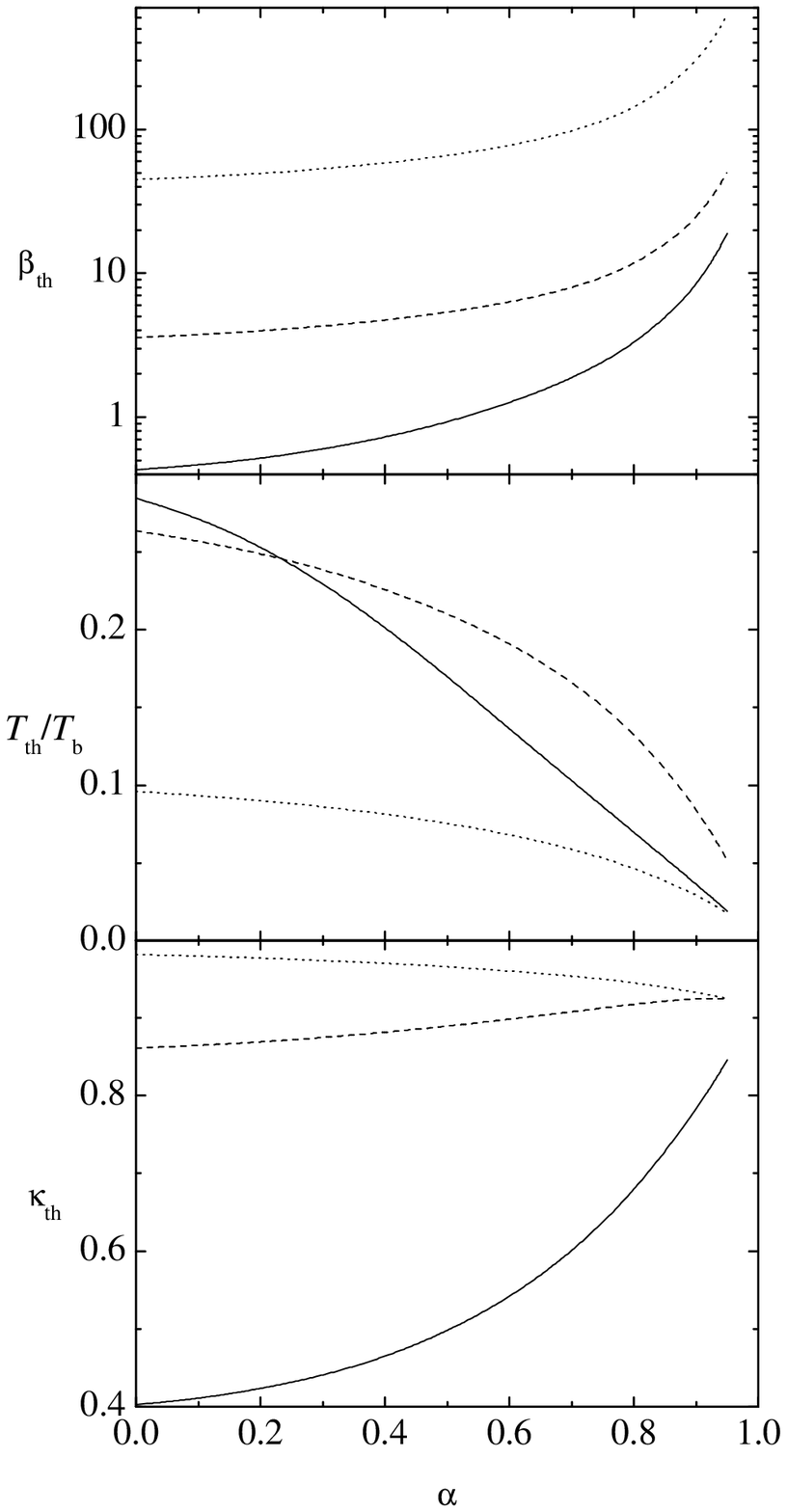}
 \caption{The top panel shows the threshold value $\beta_\kk$ versus $\alpha$ for $m/m_\bb=0.1$ (solid line), 1 (dash line), and 10 (dot line). At a given mass ratio $m/m_\bb$, the values of the temperature obtained from the Sonine and MB approximations differ less than 10\% for those points lying below the corresponding curve. The middle and bottom panels show the temperature ratio and the cumulant, respectively, on the threshold curves.\label{thresh}}
 \end{figure}
It is not feasible to know \textit{a priori} the range of validity of the  Sonine approximation. In principle, it is expected to be valid as long as the value of $|\kappa|$ remains small. But  how small must $|\kappa|$  be to trust the Sonine approximation? As comparison with numerical solutions have shown in Sec.~\ref{sec5}, the Sonine approximation gives excellent results even if $\kappa\simeq 0.1$ (cf.\ Figs.\ \ref{noise} and \ref{omega}).
Since the Sonine approximation assumes that the velocity distribution function is close to the MB distribution, a consistency test is that the values for the temperature ratio obtained from both approximations are close each other. 
As a criterion to measure the range of applicability of the  Sonine approximation, let us define a threshold value of the control parameter,  $\beta_\kk(\alpha,m/m_\bb)$, such that if $\beta<\beta_\kk(\alpha,m/m_\bb)$, then the temperature ratio obtained in the Sonine approximation differs less than 10\% from that obtained in the MB approximation.
The values of the temperature  and the cumulant at the threshold will be denoted by $T_\kk(\alpha,m/m_\bb)$ and $\kappa_\kk(\alpha,m/m_\bb)$, respectively.
Figure \ref{thresh} shows the $\alpha$-dependence of $\beta_\kk$, $T_\kk/T_\bb$, and $\kappa_\kk$ for $m/m_\bb=0.1,$ 1, and 10. 
We observe that the threshold value $\beta_\kk$ grows as the inelasticity decreases. Moreover, $\beta_\kk$  is very sensitive to the mass ratio. As the granular particles become heavier, the range of values of the control parameter $\beta$ for which the Sonine approximation is expected to be reliable widens significantly. Figure \ref{thresh} also shows that the validity criterion $\beta<\beta_\kk$ is compatible with values relatively large for the cumulant $\kappa$, especially for large $m/m_\bb$.

\begin{figure}
 \includegraphics[width=.90 \columnwidth]{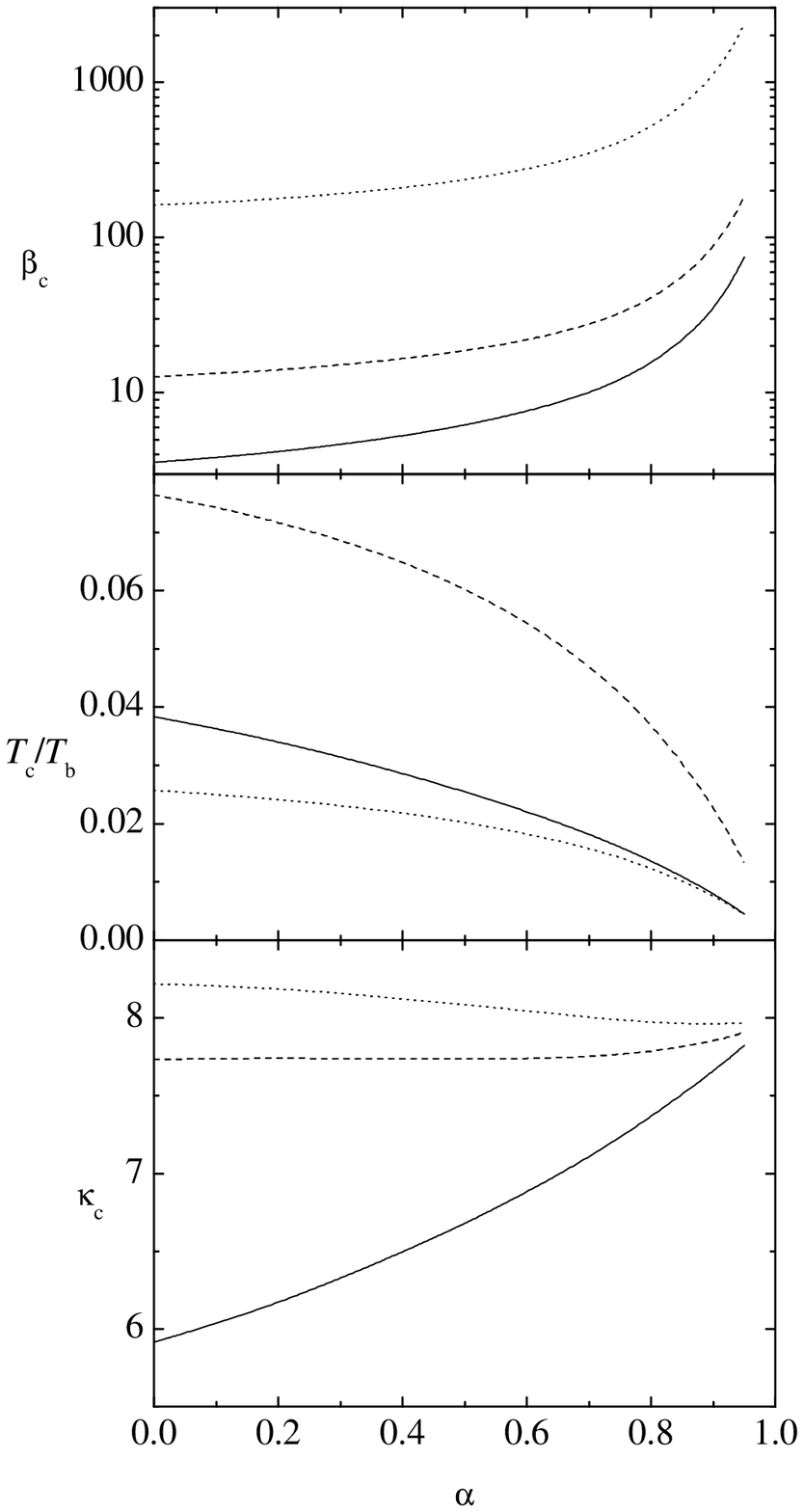}
 \caption{The top panel shows the critical value $\beta_\cc$ versus $\alpha$ for $m/m_\bb=0.1$ (solid line), 1 (dash line), and 10 (dot line). At a given mass ratio $m/m_\bb$, the Sonine approximation predicts that there does not exist a steady state if $\beta>\beta_\cc$. The middle and bottom panels show the temperature ratio and the cumulant, respectively, on the critical curves.\label{critical}}
 \end{figure}

In the MB approximation, the asymptotic behavior (\ref{b8bis}) takes place when $\beta\to\infty$, irrespective of whether this corresponds to $m/m_\bb\to \infty$ or $\omega\to \infty$. On the other hand, in the Sonine approximation the behavior (\ref{b8bis2}) reflects the limit $\beta\to \infty$ provided that $m/ m_{\bb}\to \infty$ at constant $\omega$. In other words, Eq.\ (\ref{b8bis2}) does not hold in the Sonine approximation at \textit{finite} mass ratio when $\beta\to\infty$ ($\Rightarrow \omega\to \infty\Rightarrow n/n_\bb\to \infty$), i.e.\ in the limit of a vanishing concentration of bath particles. What happens then in that limit?

The analysis of the tenth-degree equation (\ref{13.1}) shows that, as $\beta$ increases for fixed values of $\alpha<1$ and $m/m_\bb$, a critical value $\beta_\cc(\alpha,m/m_\bb)$ is reached at which the two positive real roots coalesce and no  solution with positive real $x$ exists for $\beta>\beta_c$.
 However, as $\beta\to \beta_\cc^-$, the mathematical solution of the problem tends to well-defined values $x_\cc(\alpha,m/m_\bb)$ and $\kappa_\cc(\alpha,m/m_\bb)$.
Figure \ref{critical} shows $\beta_\cc$, $T_\cc/T_\bb$, and $\kappa_\cc$ versus $\alpha$ for the same values of the mass ratio $m/m_\bb$ as in Fig.~\ref{thresh}. We observe that at given $m/m_\bb$, $\beta_\cc$ increases as the inelasticity decreases (in fact, $\lim_{\alpha\to 1^-}\beta_\cc =\infty$). Analogously, at given $\alpha$, $\beta_\cc$ increases with  the mass ratio $m/m_\bb$.
For $\beta< \beta_\cc(\alpha,m/m_\bb)$, a linear stability analysis  of Eqs.\ (\ref{c8+c9}) shows that the steady-state solution $(x,\kappa)$ is linearly stable. On the other hand, if $\beta> \beta_\cc(\alpha,m/m_\bb)$, a steady state is absent so $\kappa$ monotonically
increases and $T/T_\bb$ monotonically decreases with time. Therefore, the dynamical system (\ref{c8+c9}) presents a (first-order) \textit{phase transition} in the Sonine approximation: a stable fixed point exists for $\beta< \beta_\cc(\alpha,m/m_\bb)$ (with the ``order'' parameter $T/T_\bb\neq 0$), while no fixed point exists for $\beta>\beta_\cc(\alpha,m/m_\bb)$ (with $T/T_\bb\to 0$). 

It is difficult to assert whether the prediction of this phase transition is physically correct or it is just an artifact of the Sonine approximation employed. On the one hand, as the bath particles become more and more dilute, so the friction constant becomes smaller and smaller versus the cooling rate, it is possible that a certain critical value is reached beyond which the bath is unable to thermostat the granular fluid and, in addition, the velocity distribution of the granular particles strongly deviates from the MB form \cite{note}.
On the other hand,  the predicted critical behavior takes place clearly outside the range of validity of the Sonine approximation since the values  of the cumulant at criticality, $\kappa_\cc(\alpha,m/m_\bb)$, are too high, as seen in Fig.~\ref{critical}.  These values are so high that the Sonine description cannot be \textit{quantitatively} accurate. Whether it is \textit{qualitatively} correct when predicting the existence of a critical behavior remains an open question.

In any case, it is clear that a situation where the true $\kappa$ were small (so the Sonine approximation should then  be reliable), while the Sonine prediction of $\kappa$ were large, is not self-consistent. Thus,
one can assert that, irrespective of whether a steady state with $\beta>\beta_c$ exists or not, the true kurtosis must be quite large, indicating that the  granular distribution function must be highly distorted with respect to the MB distribution.

\section{Summary and conclusion\label{sec6}}
The intrinsic nonequilibrium nature of granular fluids appears clearly in the case of uniformly heated systems. The system considered in this paper was originally proposed by Biben et al.\ \cite{BMP02} and consists of a fluid of inelastic hard spheres (of mass $m$, diameter $\sigma$, and coefficient of restitution $\alpha$) immersed in a bath of elastic hard spheres (of mass $m_\bb$ and diameter $\sigma_\bb$) kept at equilibrium at a fixed temperature $T_\bb$. The granular particles are subjected to two competing effects: on the one hand, they experience dissipative collisions among themselves with an effective collision frequency $\nu$; on the other hand, they collide elastically against the bath particles with a characteristic rate (or friction constant) $\gamma$. The first effect tends to produce a decrease of the granular particle $T$ with an associated cooling rate $\zeta^\zero=\nu(1-\alpha^2)$, while the second effect tends to thermalize the granular fluid to the bath temperature $T_\bb$. When both mechanisms compensate each other, a steady state es reached with $T<T_\bb$ (breakdown of energy equipartition) and a nonequilibrium velocity distribution with a non-zero kurtosis $\kappa$. The stationary values of $T/T_\bb$ and $\kappa$ depend on the ratio of time scales $\beta\equiv \zeta^\zero/\gamma$, the mass ratio $m/m_\bb$, and the coefficient of restitution $\alpha$.

In order to obtain $T/T_\bb$ and $\kappa$ from the Enskog--Boltzmann equation, I have considered the (first) Sonine approximation, which is possibly the simplest approach allowing one to estimate the kurtosis of the distribution. As a consequence, $\kappa$ is expressed in terms of $T/T_\bb$, the latter quantity being given as the physical root of a tenth-degree equation. 
Explicit analytical results have been obtained in several limiting situations. In the limit of large friction constant relative to the cooling rate ($\beta\to 0$) with finite mass ratio, as well as in the colloidal limit [$m/m_\bb\to\infty$, $\beta\propto (m/m_\bb)^{-1/2}\to 0$], the breakdown of energy equipartition is weak ($1-T/T_\bb\propto \beta$) and the distribution function is close to the Maxwell--Boltzmann (MB) form ($\kappa\propto\beta$). However, if $m/m_\bb \to 0$ and $\beta\to 0$  with $(m_\bb/m)\beta\to \infty$, as well as in the white noise limit [$m/m_\bb\to \infty$, $\beta\propto(m/m_\bb)^{1/2}\to \infty$], the steady-state granular temperature is much smaller than the bath temperature and the cumulant remains finite.

In the limit of heavy granular particles ($m/m_\bb\to \infty$) the Enskog--Boltzmann operator representing the collisions of the granular particles with 
the bath particles becomes a Fokker--Planck operator. Its high energy analysis shows that the distribution function has a Gaussian tail $\ln f\approx -mv^2/2T_\bb$ for extremely large velocities ($v\gg \beta\sqrt{2T_\bb/m}$). If $\beta$ is large enough so that $T\ll T_\bb$, then there exists an intermediate window of velocities ($\sqrt{2T/m}\ll v\ll \sqrt{2T_\bb/m}$) where the distribution function has the stretched exponential form $\ln f\approx -k v^{3/2}$, the coefficient $k$ being given by Eq.\ (\ref{d8}). If the concentration and size of the granular particles are comparable to those of the bath particles, then $\beta \propto (m/m_\bb)^{1/2}\to\infty$ and, consequently, $T/T_\bb \propto (m/m_\bb)^{-1/3}\to 0$. This implies that the drift term can be neglected versus the diffusion term in the Fokker--Planck operator (for $v\ll \sqrt{2T_\bb/m}$), so the effect of collisions with the bath particles is indistinguishable from that produced by a white noise stochastic force. Thus the well-known case of a white noise thermostat is recovered from the more general case of a thermostat made of a bath of elastic hard spheres in the limit $m/m_\bb\to\infty$, while keeping fixed the concentrations and sizes of the particles. 
The fact that a granular fluid uniformly heated by a Gaussian white noise can be interpreted as a particular case of a granular fluid heated by elastic collisions with particles of an equilibrium bath highlights the physical interest of the latter system. While this system does not intend to represent faithfully the heating  mechanism by vibrations usually employed in real experiments, it embodies  most of the relevant nonequilibrium 
steady-state properties of a binary granular fluid.

For finite mass ratio $m/m_\bb$ and collision frequency ratio $\beta$ one must solve numerically the tenth-degree equation for $T/T_\bb$. Even so, this represents a formidable simplification of the problem with respect to the iterative numerical solution of the Enskog--Boltzmann equation carried out in Ref.\ \cite{BMP02}. 
Moreover, an approximate explicit solution to the tenth-degree equation can be obtained by neglecting nonlinear terms in the deviation from the MB solution. 
Comparison with the numerical data of  Ref.\ \cite{BMP02} shows an excellent performance of the results obtained from the Sonine approximation. On the other hand, the numerical solutions considered in Ref.\ \cite{BMP02} were restricted to $m/m_\bb \geq 1$ and $\beta\leq 2^{-3/2}(1+m/m_\bb)^{1/2}$ and did not explore the whole parameter space. The results presented in this paper show that, for fixed values of $\alpha$ and $m/m_\bb$, the quantitative accuracy of the Sonine approximation worsens as $\beta$ increases. For large $\beta$, the distribution function is highly distorted from the MB distribution and  so the kurtosis $\kappa$ is not small enough to make the Sonine method reliable. Quite interestingly, there exists a critical value $\beta_c(\alpha,m/m_\bb)$ such that the Sonine approximation fails to provide a steady-state solution for $\beta>\beta_c(\alpha,m/m_\bb)$. Since this phenomenon appears beyond the domain of applicability of the Sonine approximation, it might be an artifact of the approximation. Nevertheless, it is also possible that the Sonine prediction of a phase transition is qualitatively correct and the bath of elastic particles  is unable to thermalize the granular fluid if the ratio between the cooling rate and the friction constant is larger than a certain critical value \cite{note}. The investigation of this possibility will be the subject of a future work.

\begin{acknowledgments}
I wish to thank T.\ Biben for kindly providing me the data of the numerical solution of Ref.\ \cite{BMP02} and to V.\ Garz\'o for a critical reading of the manuscript.
Partial support from the Ministerio de
Ciencia y Tecnolog\'{\i}a
 (Spain) through grant No.\ BFM2001-0718 is gratefully acknowledged.
\end{acknowledgments}

\appendix*
\section{Expression for $\epsilon$}
If $x=x_0(1+\epsilon)$, $x_0(\beta)$ being given by Eq.\ (\ref{b4}), is inserted into Eq.\ (\ref{13.1}) and terms nonlinear in $\epsilon$ are neglected, one gets a linear equation for $\epsilon$ whose solution is
\beq
\epsilon=-\frac{(1-\alpha)}{32x_0^4} \frac{N(\alpha,\mu,\beta)}{D(\alpha,\mu,\beta)}\beta,
\label{A1}
\eeq
where
\begin{widetext}
\beq
N(\alpha,\mu,\beta)=
\left[16  \mu {x_0^4}+(1-2{{\alpha }^2}-16 \mu )x_0^2+2 \mu  \right]\left[64  \mu(1-\mu )  {x_0^3}+ 4  \beta 
(3+8 \mu -6 {{\mu }^2}){x_0^2}   +12 {{\beta }^2} x_0+3 {{\beta }^3}\right],
\label{A2}
\eeq
\beqa
D(\alpha,\mu,\beta)&=&
240    (1-\alpha )  \mu {x_0^{12}} -
6    \left[2(1-\alpha)(5\alpha^2+52\mu)+59 -43\alpha \right]{x_0^{10}}\nonumber\\
&&
+  \left[6(1-\alpha)(15\alpha^2+223\mu-96\mu^2+48\mu^3)+227-99{\alpha }\right]{x_0^8}\nonumber\\
&&
-  \left[2(1-\alpha)\mu(343 +46  {{\mu }}-16  {{\mu }^2})+2(1-\alpha){{\alpha }^2} (15+16  \mu -12 {{\mu
}^2})+33- \alpha \right]{x_0^6}\nonumber\\
&&
+2   (1-\alpha )  \mu   \left[42+149  \mu -34  {{\mu }^2}+2  {{\alpha }^2}  (8-5\mu)\right]{x_0^4}  
\nonumber\\
&&
-2    (1-\alpha )  {{\mu }^2}
(23+2  {{\alpha }^2}+18  \mu ){x_0^2}
+8  (1-\alpha )  {\mu }^3.
\label{A3}
\eeqa

In the limit $\beta\to 0$ with fixed $\alpha$ and $\mu$, one simply has
\beq
\epsilon\approx \left(h-\frac{1}{4}\right)\beta,
\label{A4}
\eeq
where $h$ is given by Eq.\ (\ref{c1}). 
On the other hand, if $\mu\to 0$ with fixed $\alpha$ and $\beta$, we get
\beq
\epsilon\approx \frac{3\beta^2}{8x_0^2}\frac{(1-\alpha)(1-2\alpha^2)}{10(1-\alpha)\left[16-3\alpha^2(x_0^2-1)(1-2x_0^2)\right]+(x_0^2-1)\left[127+354x_0^2-3\alpha(53+86x_0^2)\right]}
\label{A6}
\eeq
\end{widetext}
If both $\beta$ and $\mu$ go to 0, Eq.\ (\ref{A1}) reduces to
\beq
\epsilon\approx \frac{1-2\alpha^2}{1280}(3\beta+4\mu)\beta.
\label{A7}
\eeq
In particular, in the colloidal limit ($\beta\propto \mu^{1/2}\to 0$), one gets
\beq
\epsilon\approx \frac{3(1-2\alpha^2)}{1280}\beta^2.
\label{A5}
\eeq
The same result is obtained by making $x_0\to 1$ in Eq.\ (\ref{A6}).

Finally, in the white noise limit ($\beta\propto \mu^{-1/2}\to \infty$), Eq.\ (\ref{A1}) yields
\beq
\epsilon\approx \frac{1}{4}\frac{(1-\alpha)(1-2\alpha^2)}{16+(43+10\alpha^2)(1-\alpha)}.
\label{A8}
\eeq
Comparison with Eq.\ (\ref{c3}) shows that Eq.\ (\ref{A8}) is close to $\epsilon\approx (2\lambda-1)/3$.

\end{document}